\documentclass[%
 reprint,
 amsmath,amssymb,
 aps,
]{revtex4-1}

\pdfoutput=1

\usepackage{graphicx}
\usepackage{dcolumn}
\usepackage{bm}

\begin{document}

\preprint{APS/123-QED}

\title{
Simple Empirical Model\\for Identifying Rheological Properties of Soft Biological Tissues
}

\author{Yo Kobayashi}
\email{you-k@fuji.waseda.jp}%
\affiliation{Faculty of Science and Engineering /Research Institute of Science and Engineering, Waseda University, Tokyo, Japan \\ and JST-PRESTO}%

\author{Mariko Tsukune}%
\author{Tomoyuki Miyashita}%
\author{Masakatsu G. Fujie}%
\affiliation{%
 Faculty of Science and Engineering, Waseda University, Tokyo, Japan 
}%






\date{\today}

\begin{abstract}

Understanding the rheological properties of soft biological tissue is a key issue for mechanical systems used in the healthcare field. We propose a simple empirical model using Fractional Dynamics and Exponential Nonlinearity (FDEN) to identify the rheological properties of soft biological tissue. The model is derived from detailed material measurements using samples isolated from porcine liver. We conducted dynamic viscoelastic and creep tests on liver samples using a rheometer. The experimental results indicated that biological tissue has specific properties: i) power law increases in storage elastic modulus and loss elastic modulus with the same slope; ii) power law gain decrease and constant phase delay in the frequency domain over two decades; iii) log-log scale linearity between time and strain relationships under constant force; and iv) linear and log scale linearity between strain and stress relationships. Our simple FDEN model uses only three dependent parameters and represents the specific properties of soft biological tissue.

\end{abstract}

\maketitle


\section{\label{intro}Introduction}

\subsection{\label{backg}Background}

Understanding the rheology ---the study of materials with both solid and fluid characteristics in which the response to strain under applied stress is evaluated--- of biological tissues is a key issue for current research in the human healthcare field. Rheology is relevant to many technological applications, ranging from biological science (e.g. medicine, sports, biology, biomechanics) to engineering (e.g., robotics, mechatronics, material mechanics, control theory, computational mechanics, information technology). Recently, the healthcare field has realized the benefit of using intelligent machines (such as robots) that can physically interact with human. As a byproduct, the physical information measured by the machines can also be used for cyber system construction (such as machine learning). Understanding the physical phenomena underlying the mechanical properties of human tissue has a great impact on bio-science and engineering. This knowledge will lead to further development of machines and systems in the healthcare field.

Modeling of soft tissue rheological properties is a core technology for developing various healthcare machines and systems to assist human activity. For example, a mathematical model of target objects (human, organ, tissue, etc.) is required for mechanical design, motion planning, information processing, and machine/system control. These research and development areas require fundamental equations that are limited to the essential properties of the macroscopic behavior of the target matter (i.e., micro-scale modeling is not necessary). In short, the development of fundamental macroscopic models of the properties of biological matter are a key research issue pertinent to healthcare machines and systems designed for humans, organs, and tissues.

In spite of their scientific and technological importance, mainly because they are difficult to model, very little knowledge has been established regarding the rheological properties of soft biological tissues. The problem is due to a lack of established methods for sensing, parameterizing, and information processing of rheological properties of soft biological tissue. The rheological properties of soft biological tissue cannot be directly modeled in the same manner as synthetic matter. In general, an ordinary Linear Differential Equation (LDE) is used to model the target object. In other words, the terms of the equation for rheological properties have been generally modeled using both 'Linear' and 'Integer order' differential equations explicitly or implicitly. However, the properties of soft biological tissues are different from synthetic matter and the LDE model tends to be inaccurate for data derived from experiments using soft biological tissue. Therefore, we aimed to develop a simplified fundamental macroscopic model of the rheological properties of soft biological tissue that provides a good fit for experimental data.

\subsection{\label{goal}Goal and Motivation}
The goal of this study is to establish a universal fundamental model to represent the macroscopic rheological properties of soft biological tissue, as well as a measurement method for these properties. Many researchers have reported that the rheological properties of soft biological tissues have distinct properties in comparison to industrial synthetic materials, such as metals. For example, researchers reported that biological tissues have viscoelastic properties. Researchers have also reported that soft biological tissues exhibit a very nonlinear relationship between strain and stress. In this article, we describe the complex viscoelastic and nonlinear properties of soft biological tissue as 'rheological properties'.
The motivation of this study is to propose a 'simple model' that accurately represents the specific rheological properties of soft biological tissue. The model should be strongly correlated with experimental data derived from actual biological tissue. A 'simple model' means that the model should utilize the minimum number of parameters, yielding a mathematical equation that is easy to understand and implement. Use of a simple model is essential for robust identification and discrimination of tissues using stress and strain information.

\subsection{\label{relat}Related research}
Numerous studies have dealt with both the nonlinearity or/and viscoelasticity of biological tissue 
 \cite{fung1981biomechanics,Miller2002,Sarver2003,Brands2004,
 chui2004combined,kerdok2006effects,Sedef,Samur2007,Tanaka2008,
 gao2010constitutive,SadeghiNaini2011,Darvish2001,kim2003characterization,
 schwartz2005modelling,Kim2005,liu2007development,
 famaey2008soft,Lu2010,Sims2010,Marchesseau2010,Ahn2010,
 Basafa2011,Basafa2011}. 
 An Ordinary Linear Differential Equation (LDE) is generally used to model viscoelastic properties (e.g. Voigt/Maxwell/Kelvin model). However, LDE models do not fit data from biological tissues well. Furthermore, a large number of parameters are used in LDE to increase model accuracy (e.g. generalized Voigt　/　Maxwell model), and these models only represent linear relationships between stress and strain. Hyperelastic models (e.g. Ogden, Moonine-Revlin model) are generally used to represent stress-strain nonlinearity, although the number of parameters in hyperelastic models also tends to be large. Moreover, these models are time-independent and does not represent dynamic (viscoelastic) properties. Models that neglect viscoelasticity and/or nonlinearity result in variability and incongruous analysis of the rheological properties of soft biological tissue. Specifically, parameter robustness decreases because the parameters readily change due to the duration of stress application and/or the magnitude of strain.

The equations in the some related works 
 \cite{Darvish2001,kim2003characterization,
 schwartz2005modelling,Kim2005,liu2007development,
 famaey2008soft,Lu2010,Sims2010,Marchesseau2010,Ahn2010,
 Basafa2011,Basafa2011}
have dealt with both viscoelasticity and nonlinearity. However, these models tend to become overly complex and involve an excess number of material parameters to represent these properties. We believe that a preferred model should have a small number of parameters that are strongly correlated with the experimental data. Existing models with numerous parameters ---such as those combining hyperelastic models with viscoelastic models--– are unsuitable for identifying model parameters. The use of a large number of parameters leads to a risk overfitting of the parameter identification and ill-posedness of inverse problems. The large parameter number also increases computational costs.
A model based on a simple equation with few parameters that is highly correlated with experimental data from soft biological tissues does not currently exist. 

Therefore, we have conducted studies aimed at developing a model with these characteristics 
\cite{Kobayashi2005,Kobayashi2009,Kobayashi2012,kobayashi2012viscoelastic,
Kobayashi2012Enhanced,Tsukune2014,Kobayashi2012Soft,okamura2014study}
. The model is derived from comprehensive material data obtained from in vitro measurements of porcine liver   \cite{Kobayashi2005,Kobayashi2009,Kobayashi2012,kobayashi2012viscoelastic}
. The model was also validated using in vitro breast tissue (fibroglandur tissue, fat, muscle) 
\cite{Kobayashi2012Enhanced,Tsukune2014} and partially evaluated in muscle tissue \cite{Kobayashi2012Soft,okamura2014study}
. The model combines a fractional differential equation with a polynomial expression for stress-strain nonlinearity, which consists of four parameters \cite{Kobayashi2005,Kobayashi2009,Kobayashi2012,kobayashi2012viscoelastic,
Kobayashi2012Enhanced,Tsukune2014}. However, the two parameters in the model ---both parameters representing nonlinear properties--- correlate and interfere with one another. In addition, the parameter identification from the experimental data of these two parameters is complex; specifically, global searching and optimization is necessary.

\subsection{\label{object}Objectives}
\begin{figure*}
\includegraphics[width=17cm]{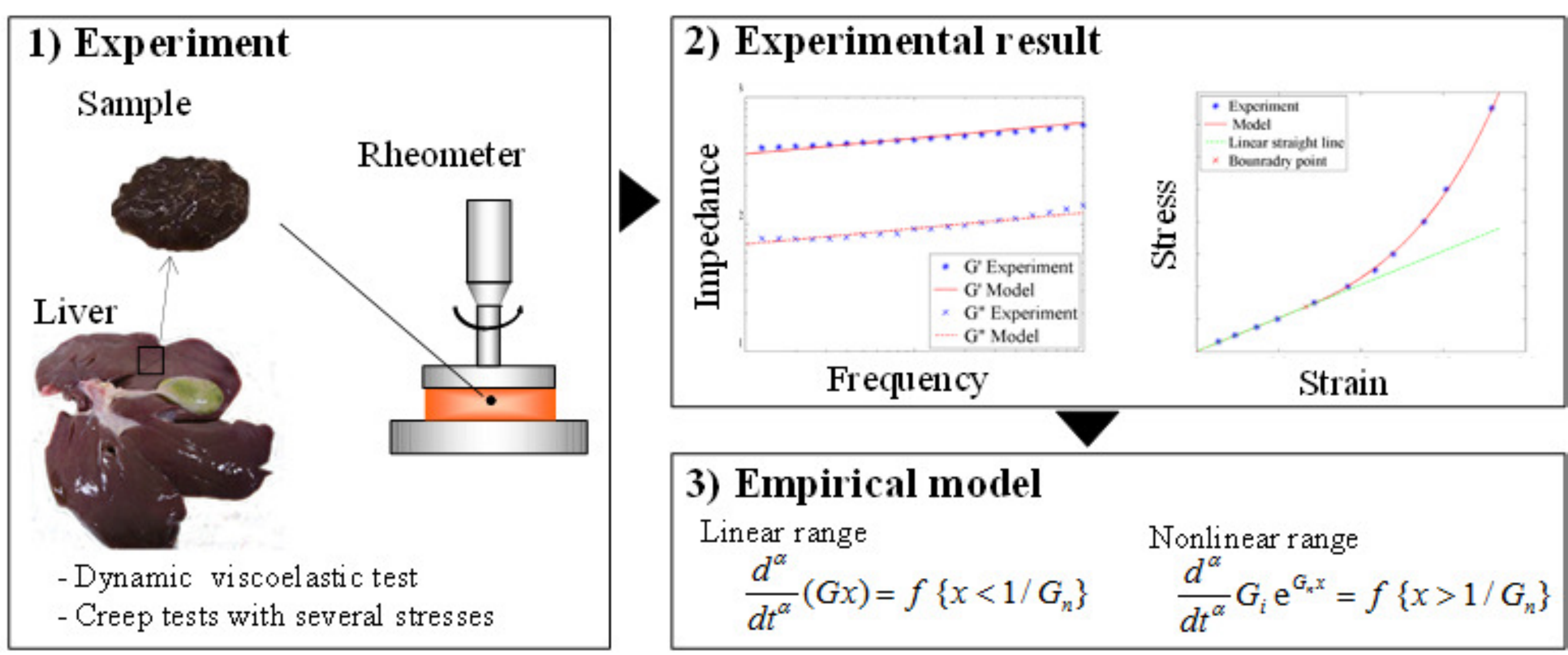}
\caption{\label{fig:concept}The overview of this article. The model is derived from detailed material measurements using actual biological tissue. Specifically, we used samples isolated from various porcine livers. We used a rheometer to measure the liver samples in which the rheometer can dynamically control and measure stress and strain applied to the sample. Rheometers are commonly used to measure the fundamental physical properties of food samples. For example, we conducted a dynamic viscoelastic test and the creep test with several stresses to derive and evaluate our empirical model. Individual differences between liver samples –-physical properties of biological tissues differ between individual samples-– were represented by the values of model parameters.} 
\end{figure*}

The objective of this article is to propose a simple model that represents the rheological properties ---meaning, viscoelastic and nonlinear properties---  of soft biological tissues. Specifically, we propose a simple model, using only three dependent parameters, incorporating fractional dynamics and exponential nonlinearity to identify rheological properties. The advantage of our model is that it is strongly correlated with various experimental data and uses a small number of parameters, thereby rendering it suitable for parameter identification and inverse analysis. 
Figure \ref{fig:concept}  shows an overview of this article. The model is derived from detailed material measurements using actual biological tissue. Specifically, we used samples isolated from various porcine livers. We selected liver samples because liver is a relatively simple tissue with low anisotropy when compared with other biological organs and tissues. We used a rheometer to measure the liver samples, as the rheometer can dynamically control and measure stress and strain applied to the sample. We conducted a dynamic viscoelastic test and creep test to derive and evaluate the model. Individual differences between liver samples –-physical properties of biological tissues differ between individual samples-– were represented by the values of model parameters. 

\section{\label{mater}Materials and Methods }

In this section, we explain how we measured and modeled the rheological properties ---nonlinear and viscoelastic properties--- of the samples. First, we introduce our rheological model scheme. We then explain the study materials and measurement procedures.

\subsection{\label{propo}Proposed model }

The rheological model in this study relies on experimental data obtained from biological tissues. We first denote the model equations (\ref{eq:LinearFrac}) 
(\ref{eq:NonlinearFrac}) to enhance the readability of this article. The proposed rheological model utilizes Fractional Dynamics and Exponential Nonlinearity (FDEN). The equations are as follows:

\begin{subequations}
\begin{eqnarray}
 \frac{{{d^\alpha }}}{{d{t^\alpha }}}(Gx) = f \quad \ \left\{ x < {x_b}\right\} \
\label{eq:LinearFrac}. 
\\
\frac{{{d^\alpha }}}{{d{t^\alpha }}}({G_i}{e^{{G_{n}}x}}) = f  \quad \ \{ x > {x_b}\} \
\label{eq:NonlinearFrac}. 
\end{eqnarray}
\label{eq:FDEN}
\end{subequations}

where \textit{x} is strain (torsional strain), \textit{f} is stress (torsional stress) and \textit{t} is time, as variables; \textit{$\alpha$} is a non-integer derivative order representing the index of viscoelasticity, \textit{G} is linear viscoelastic stiffness, \textit{$G_{n}$} is nonlinear viscoelastic stiffness (\textit{$G_{i}$} is dependent parameter), and \textit{$x_{b}$} is the boundary strain in which the characteristics change to nonlinearity, as the parameters of the model. \textit{e} is Napier's constant. Each parameter should fulfill the following relationship concerning the connectivity between linear (\ref{eq:LinearFrac}) and nonlinear
(\ref{eq:NonlinearFrac}) equations –--the exponential curve 
(\ref{eq:NonlinearFrac}) is a tangent to the straight line(\ref{eq:LinearFrac})–--.  

\begin{eqnarray}
{x_b} = \frac{1}{{{G_n}}}, \quad \ {{G_i} = \frac{G}{{{G_n}e}}}
\label{eq:parameter1}. 
\end{eqnarray}

The total number of parameters in the model is three (\textit{$\alpha$}, \textit{G} and \textit{$G_n$} as representative parameters) according to the relationships in (\ref{eq:parameter1}). The details of the experimental methods and derivation process of the model from the experimental data are described in the next sections.

\subsection{\label{mater}Materials and conditions }

\begin{figure}[b]
\includegraphics[width=8.5cm]{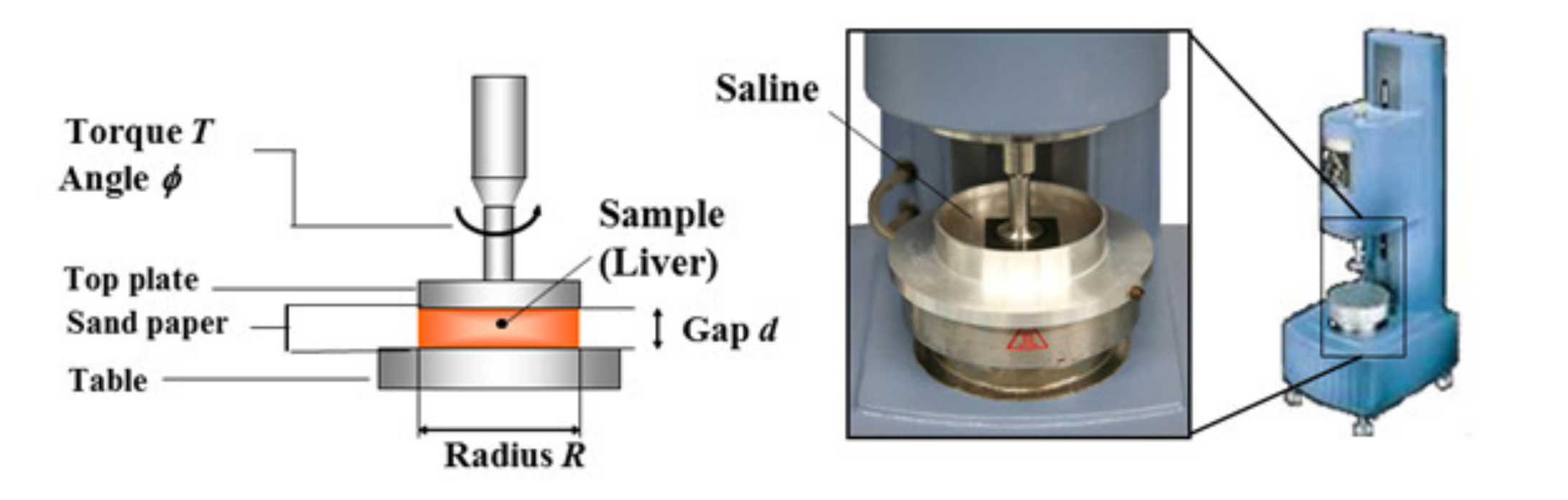}
\caption{\label{fig:rheometer} The details of the measuring components. We used porcine liver as sample of this study. We used a rheometer to measure the stress loaded on the sample and sample strain. The liver sample was cut into slices and placed on a measurement table. The samples were soaked in a saline solution at 35 °C during each test. Sandpaper was attached to the top plate and the measurement table to prevent sliding. d and R are the length and radius of the cylinder equation (A-1) and (A-2). R was 20 [mm] and d was 5 [mm] in the experimental setup of this paper.} 
\end{figure}

Figure  \ref{fig:rheometer} shows the details of the measuring components. We used porcine liver in the present study because porcine abdominal organs have properties similar to those of humans, for example, the porcine abdominal organs are widely used in laparoscopic surgery training for novice surgeons. We chose to measure the properties of liver samples because we thought that liver would be relatively easy to model –--liver consists of homogeneous and isotropic tissue–--. We used cryogenically preserved liver samples (4°C on ice) that were taken within 24 hours post-mortem and that did not include membranes or large blood vessels. Specimens were not frozen at any time during the procedure. 
We used a rheometer (AR550 or AR-G2; TA Instruments, New Castle, DE) to measure the stress loaded on the sample and sample strain. The shear stress rheometer was selected because the shear test must be independent with the change of cross-sectional area in the stress calculation. In addition, the effect of gravity could be disregarded. From these measurements, the conventional shear strain \textit{x} and conventional shear stress \textit{f} were calculated, respectively. The details of the calculation are described in the  Appendix.\ref{ShearStressAndStrain}. 
The liver sample was cut into slices (diameter 20 mm, height 5 mm) and placed on a measurement table. The samples were soaked in saline solution at 35°C during each test. Sandpaper (P80 grain size) was attached to the top plate and the measurement table to prevent sliding.

\subsection{\label{proce} Procedures}

\subsubsection{Initializing procedures}
After the saline solution reached the target temperature, the gap was zeroed to the surface of the saucer. The saline solution was stable and there was no reflux flow. Each tissue sample was placed on a measurement table, and its thickness (=gap) was determined. The sample thickness was defined as the distance between the surface of the saucer and the surface of the parallel plate (part of the measuring device) at the time that the normal stress resulting from contact between the parallel plate and the sample reached 0.1 N. To engage the sample and parallel plate, preloading for 3 minutes and unloading for 3 minutes was performed thrice under a load shear stress of 750 Pa. 
The following series of experiments were conducted for each sample, after the above initializing procedures. 

\subsubsection{Dynamic viscoelastic test}
Sine-wave stress from 0.1 to 10 rad/s, providing 3\% strain amplitude, was applied to the sample. The gain and phase delay of each frequency were measured. Then, gain (from stress to strain), phase delay, and mechanical complex impedance of the sample in each frequency were calculated. As shown in following experimental results (Fig.\ref{fig:NMF}), 3\% (= 0.03) strain amplitude is in the range where liver tissue exhibits linear responses. The effect of mass (inertia) and shear viscosity from the external normal saline solution could be disregarded at frequencies less than 10 rad/s. Data were collected from 6 liver samples. 

\subsubsection{Creep test and nonlinear measurement}
Torsional creep test was performed after dynamic viscoelastic test. The creep test, in which step responses to strain are observed under constant stress, was repeatedly performed, applying several stresses on the sample. Time series of strain data were measured during each experiment. The shear stress load ranged from 25 to 750 Pa, and the time series of strain data were recorded for 180 seconds at each stress level. Each test was performed at intervals of 180 seconds. The load shear stress during each interval was 0 Pa. The reference strain was set to 0 at each creep test to account for residual stress and strain. We ignored the data obtained from 0 to 1 s because of vibrations during the early transient stage. The details of this area are presented in our previous article \cite{Kobayashi2012,kobayashi2012viscoelastic}. Data were collected from 64 liver samples. 

\section{\label{resul}Results and Modeling}

\subsection{\label{mecha} Mechanical complex impedance}

Here, mechanical complex impedance is defined as follows:

\begin{eqnarray}
{ {G^*}  = G' + jG''}
\label{eq:defMI}. 
\end{eqnarray}
where \textit{j} is unit imaginary number, \textit{$G^*$} is complex mechanical impedance, \textit{G'} is storage elastic modulus and \textit{G''} is loss elastic modulus. 

\begin{figure}[b]
\includegraphics[width=8.5cm]{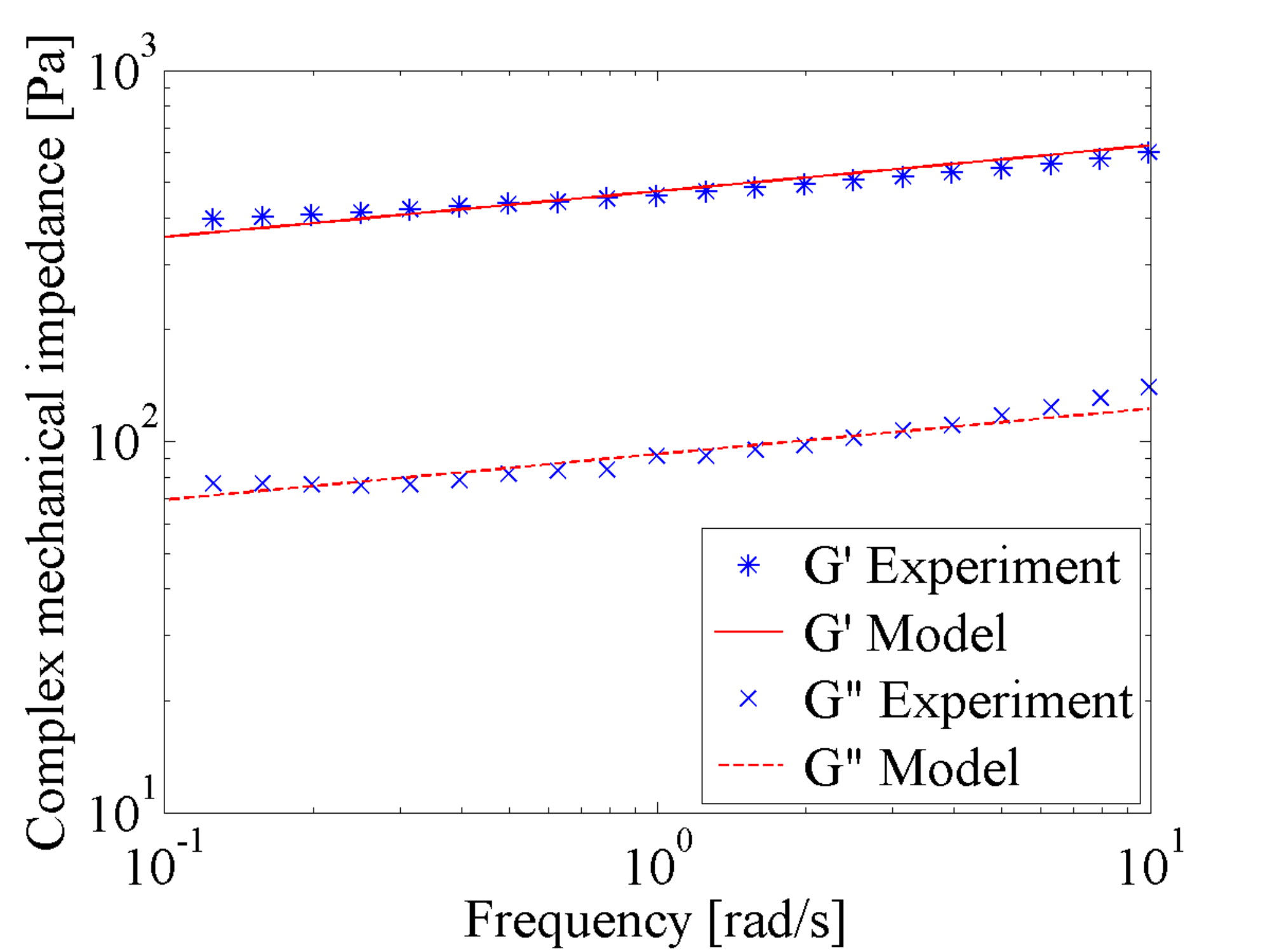}
\caption{\label{fig:MI} The typical experimental result of a dynamic viscoelastic test –--in this figure, mechanical complex impedance–-- of a sample. Blue plot is experimental result of storage elastic modulus \textit{$G'$}. Red plot is experimental result of loss elastic modulus \textit{G’’}. Both the storage elastic modulus \textit{$G'$} and the loss elastic modulus \textit{G''} increased as the frequency increased. Both storage elastic modulus \textit{$G'$} and loss elastic modulus \textit{G''} exhibit a power law form over two decades. Furthermore, there is linearity in the log-log diagram in the change of \textit{$G'$} and \textit{G''}. In addition, the slopes of \textit{$G'$} and \textit{G''} in the log-log diagram are nearly identical (approximately 0.125= 1/8). The \textit{$G'$} of our model is blue line. \textit{G''} of our model is red line. The \textit{$G'$} and \textit{G''} of our model, which parameters fit the typical experimental results, indicate that our model and the experimental results are highly correlated.
} 
\end{figure}

Typical experimental results of a dynamic viscoelastic test –-in this section, mechanical complex impedance-– of a sample are shown in Fig.\ref{fig:MI}. All liver samples exhibit the same trend as the typical sample; data trends are the same, however, model fit data and parameters are different.

Both the storage elastic modulus \textit{G'}  and the loss elastic modulus \textit{G''} increased with the frequency $\omega$. We found that both storage elastic modulus \textit{G'}  and loss elastic modulus \textit{G''} exhibit a power law form over two decades. Furthermore, there is linearity in the log-log diagram in the change of \textit{G'}  and \textit{G''}, while the slopes of \textit{G'} and \textit{G''} in the log-log diagram are nearly identical (approximately 1/8=0.125).

The mechanical complex impedance of our model has the same characteristics as the experimental results, i.e. power law form of \textit{G'}  and \textit{G''}, and same slopes of \textit{G'}  and \textit{G''}. The expansion of the equation to describe the explanation of the above characteristics is as follows. Our model is represented as equation (\ref{eq:LinearModelAgain}) –--the same equation as (\ref{eq:LinearFrac}) is described for readability–-- because the dynamic viscoelastic tests were conducted on a linear range in the stress and strain relationship.

\begin{eqnarray}
\frac{{{d^\alpha }}}{{d{t^\alpha }}}(Gx) = f
\label{eq:LinearModelAgain}. 
\end{eqnarray}

Because equation (\ref{eq:LinearModelAgain}) takes the form of a frequency transfer function, the complex shear modulus \textit{$G^*$} can be expressed in terms of the Laplace operator s as follows:

\begin{eqnarray}
{G^*(s)}=\frac{{F(s)}}{{X(s)}} = G\,{s^\alpha }
\label{eq:Laplace}. 
\end{eqnarray}

Equation (\ref{eq:MIModel}) derived from the mechanical complex impedance of (\ref{eq:Laplace}).

\begin{eqnarray}
{G^*} = G\,{(j\omega )^\alpha }
\label{eq:MIModel}. 
\end{eqnarray}

Where $\omega$ is the frequency. 
Equation (\ref{eq:MIModel}) expands to (\ref{eq:Gp}) and (\ref{eq:Gpp}) with separation of the real and imaginary parts of  (\ref{eq:MIModel}).

\begin{subequations}
\begin{eqnarray}
G' = G{'_0}{\omega ^\alpha }
\label{eq:Gp}. 
\\
G'' = G'{'_0}{\omega ^\alpha }
\label{eq:Gpp} 
\end{eqnarray}
\label{eq:GpandGpp}
\end{subequations}

Where \textit{$G'_{0}$} is a constant parameter that represents storage elastic modulus and  \textit{$G''_{0}$} is a constant parameter that represents loss elastic modulus. These parameters have the following relationship:

\begin{subequations}
\begin{eqnarray}
G = \sqrt {G{{_0^{'}}^2} + G{{_0^{''}}^2}}
\label{eq:Relation_Gp_Gpp}. 
\\
G_0^{'} = G\cos (\frac{\pi }{2}\alpha )
\label{eq:Gp2}. 
\\
G_0^{''} = G\sin (\frac{\pi }{2}\alpha )
\label{eq:Gpp2}. 
\end{eqnarray}
\end{subequations}

Euation (\ref{eq:logGp}) and (\ref{eq:logGpp}) were derived from (\ref{eq:Gp}) and (\ref{eq:Gpp}) by log-log transformation.

\begin{subequations}
\begin{eqnarray}
\log G' = \alpha \log \omega + \log G{'_0}
\label{eq:logGp}. 
\\
\log G'' = \alpha \log \omega  + \log G'{'_0}
\label{eq:logGpp}. 
\end{eqnarray}
\end{subequations}

Thus, our model equation represents the trend in the experimental results, i.e. linearity on log-log diagram. 

The parameters (\textit{G} and \textit{$\alpha$}) of equations (\ref{eq:logGp}) and (\ref{eq:logGpp}) were identified by fitting the experimental results for each sample. We used the Extended Kalman Filter (EKF) algorithm to identify the parameters (ref. Appendix C in detail) because equations (\ref{eq:logGp}) and (\ref{eq:logGpp}) are nonlinear simultaneous equations –--both equations include parameters  (\textit{G} and \textit{$\alpha$})---. The \textit{G'} and \textit{G''} in our model, which fit typical experimental results, are shown in Fig. \ref{fig:MI}, showing that our model and the experimental results are strongly correlated. It should be noted that the order of derivative \textit{$\alpha$} is not an integer; it is approximately 0.125 (= 1/8). Our model also fits all liver samples well. The coefficient of determination \textit{$R^2$} between our model and the experimental data from the frequency series of \textit{G'} and \textit{G''} in all samples is about 90\%. Tables \ref{tab:AccuracyEvaluation} and \ref{tab:FundamentalStatistics} list the model accuracy evaluation and fundamental statistics of model parameters.

\subsection{\label{bode} Bode diagram}

\begin{figure}[b]
\includegraphics[width=8.5cm]{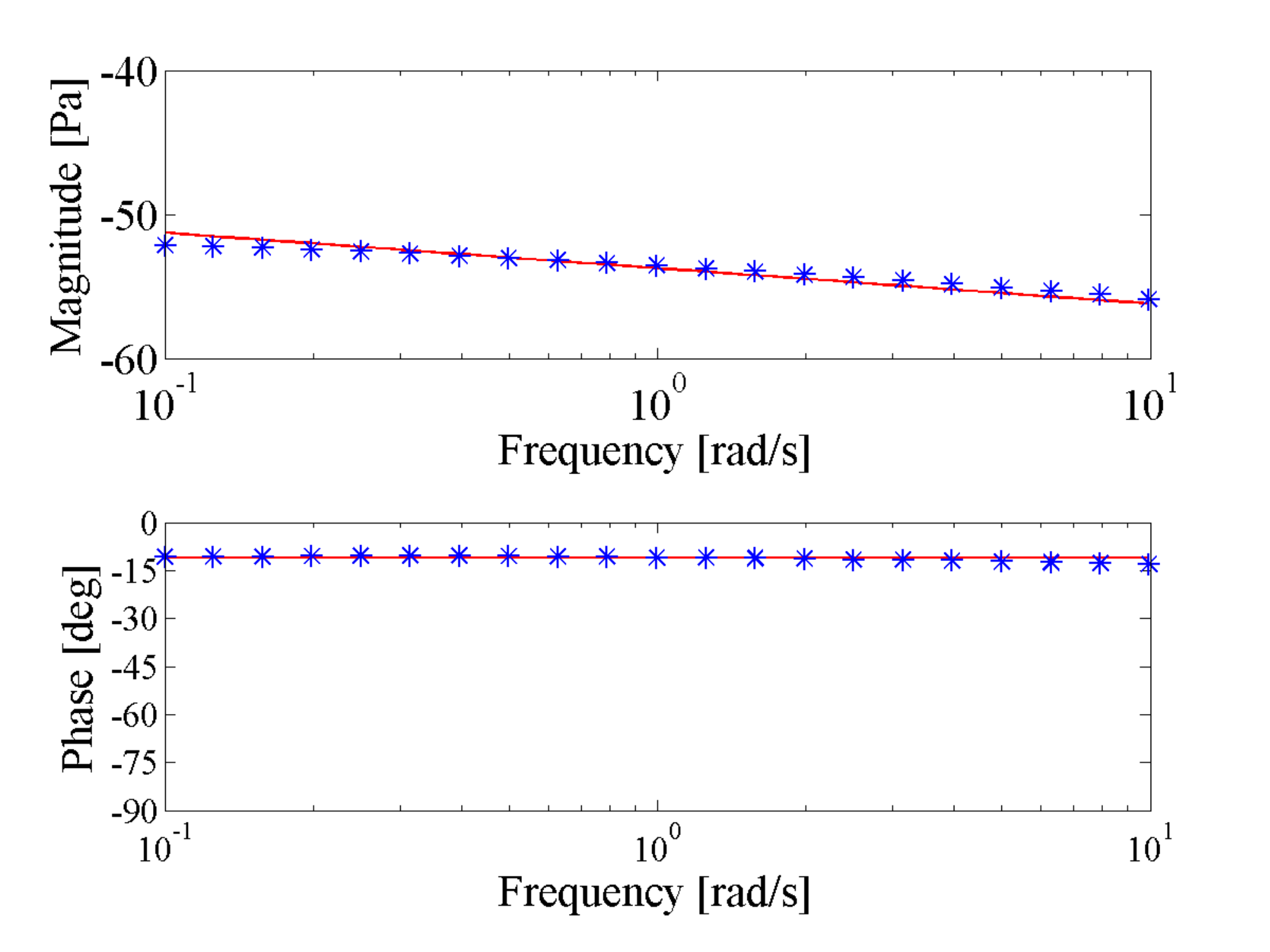}
\caption{\label{fig:BodeF} The typical experimental result of a dynamic viscoelastic test –--Bode diagram (gain diagram and phase diagram)--- of a sample. Blue plots in gain diagram and phase diagram show the experimental result. Gain –--multiplicative inverse of \textit{$G^*$}--– decreased as the frequency increased. Gain assumes a power law form over two decades. Indicating that there is linearity between frequency and gain in the log-log diagram. We also found that phase delay remained constant over two decades. The gain and phase results of our model are red lines. Our model, which the parameters fit to the typical experimental results shows that our model and the experimental results are highly correlated.} 
\end{figure}

Typical experimental results of dynamic viscoelastic test –--in this section, gain diagram and phase diagram –-- of a sample are shown in Fig.\ref{fig:BodeF}. Figure \ref{fig:BodeF} is a plot of the same data presented in Fig.\ref{fig:MI}, the only difference being the expression of dynamic viscoelastic tests data. Gain –--multiplicative inverse of \textit{$G^*$}--– decreased as frequency increased. We found that gain assumes a power law form over two decades, indicating that there is linearity between frequency and gain in the log-log diagram. We also found that phase delay remained constant over two decades. 

The bode diagram of our model has the same characteristics as the experimental result, namely power law form of gain and constant phase delay. The expansion of the equation to describe the explanation of the above characteristics is as follows. The Laplace operator of the bode diagram is as follows:

\begin{eqnarray}
\frac{{X(s)}}{{F(s)}} = \frac{1}{{G{s^\alpha }}}\, = \frac{1}{{G{\kern 1pt} (j\omega ){}^\alpha }}
\label{eq:Laplace2} 
\end{eqnarray}

The model equations of gain, (\ref{eq:Gain}) and (\ref{eq:GainLog}), are simply calculated from (\ref{eq:Laplace2}) as follows. Equation (\ref{eq:GainLog}) is derived from log-log transformation of (\ref{eq:Gain}): 

\begin{subequations}
\begin{eqnarray}
Gain= \left| {\frac{1}{{G(j\omega )^\alpha}}} \right| = \frac{1}{{G{\omega ^\alpha}}}
\label{eq:Gain}. 
\\
\log (Gain)= - \log G - \alpha \log \omega
\label{eq:GainLog}. 
\end{eqnarray}
\end{subequations}

In addition, the model equation of phase delay is derived from (\ref{eq:Gp}) and (\ref{eq:Gpp}).

\begin{eqnarray}
Phase = \arg ( {\,\frac{{G'}}{{G''}}\,})= \arg \,({\frac{1}{{G {{(j\omega )}^\alpha }}}} ) =  - \frac{\pi }{2}\alpha 
\label{eq:Phase}. 
\end{eqnarray}

Thus, our model equation represents the trend observed in the experimental results. 

We calculated the gain and phase of our model via identification of the parameter of mechanical complex impedance for each sample because parameters were same. The gain and phase results from our model, which the parameters fit to the typical experimental results, are shown in Fig.\ref{fig:BodeF}, which shows that our model and the experimental results are strongly correlated. Our model fit all liver samples well. Tables \ref{tab:AccuracyEvaluation} and \ref{tab:FundamentalStatistics} list the model accuracy evaluation and fundamental statistics of model parameters.

\subsection{\label{creep} Creep test (Step response)}

\begin{figure}[b]
\includegraphics[width=8.5cm]{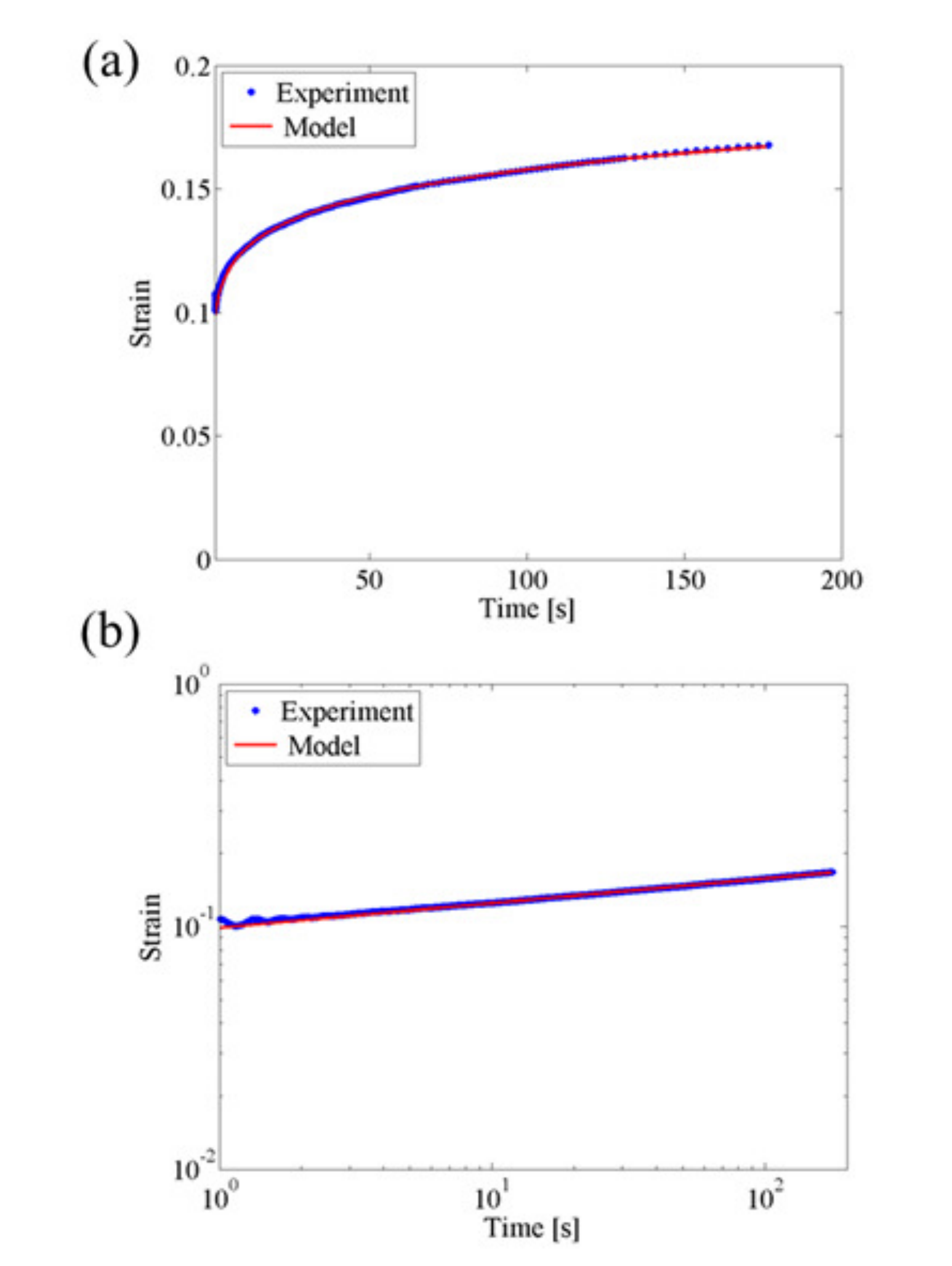}
\caption{\label{fig:CreepF} (a) A typical example of experimental results of a creep test –--the creep response obtained by assuming the input step stress--–. (b) shows a log-log diagram of the same data described in (a). The Blue plot in (a) and (b) are experimental results. The strain of the sample increased over a time interval of 180 s. We found the time series data of the creep response exhibited a power law form over two decades. This indicates that there is linearity between time and strain in the log-log diagram. The time series data of our model are red lines. Our model equation represent the trend observed in the experimental results. These figure shows our model and the experimental results are highly correlated.
} 
\end{figure}

Typical example of experimental results of a creep test –--the creep response obtained by assuming the input step-stress--– is shown in Fig.\ref{fig:CreepF}(a). The strain of liver samples increased over a time interval of 180 s. Figure \ref{fig:CreepF}(b) shows a log-log diagram of the same data described in Fig.\ref{fig:CreepF}(a). We found the time series data of the creep response exhibited a power law form over two decades. This indicates that there is linearity between time and strain in the log-log diagram (Fig.\ref{fig:CreepF}(b)).
A model equation of strain in the creep test can be calculated. We assumed that equation (\ref{eq:LinearModelAgain}) is valid for a single creep test, while nonlinearity was evaluated by a series of creep tests under several applied stresses. Specifically, equation (\ref{eq:LinearModelAgain}) becomes (\ref{eq:Creep}) if (\ref{eq:LinearModelAgain}) is solved for the conditions of the creep test. Here, the applied stress is constant \textit{$f_c$}. Equation (\ref{eq:CreepLog}) is derived from log-log transformation of (\ref{eq:Creep}).

\begin{subequations}
\begin{eqnarray}
x = \frac{{{f_c}}}{{G\Gamma (1 + r)}}{t^\alpha }( = {x_c}{t^\alpha })
\label{eq:Creep}. 
\\
\log x = \alpha \log t + \log {x_c}
\label{eq:CreepLog}. 
\end{eqnarray}
\end{subequations}

where \textit{x} is strain and \textit{t} is time; \textit{$f_c$} is constant stress; \textit{G} is proportional factor representing stiffness at each strain, as parameter \textit{$\Gamma()$} is the gamma function. \textit{$x_c$} is the coefficient determining the strain value as a parameter, which is defined as follows:

\begin{eqnarray}
{x_c} = \frac{{{f_c}}}{{G\Gamma (1 + \alpha )}}
\label{eq:xc}. 
\end{eqnarray}

In this case, the Riemann-Liouville definition (\ref{eq:FC}) –--but not only this definition–-- was used to solve the fractional integration of (\ref{eq:LinearModelAgain}). 

\begin{eqnarray}
{D^{ - \alpha }}f(t) = \frac{1}{{\Gamma (\alpha )}}{\int_{\,0}^{\,t} {(t - \xi )} ^{\alpha  - 1}}\,f(\xi )\,d\xi
\label{eq:FC}. 
\end{eqnarray}

Thus, our model equation represents the trend observed in the experimental results. The parameters (\textit{$\alpha$}, \textit{$x_c$}) of equation (\ref{eq:CreepLog}) were identified by fitting the experimental results in the log-log domain. We used the LSM algorithm to identify the parameters of equation (\ref{eq:CreepLog}) –--linear regression--– for each sample. We calculated the other independent parameter \textit{G} via equation (\ref{eq:xc}). The time series displacement data from our model, the parameters of which fit the typical experimental results, are shown in Figs.\ref{fig:CreepF}(a) and (b). Figs.\ref{fig:CreepF}(a) and (b) show that our model and the experimental results are strongly correlated. Our model fit all liver samples well. The coefficient of determination  \textit{$R^2$} between our model and the experimental data from the time series of displacement in all samples at all stresses exceeded 99\%.Tables \ref{tab:AccuracyEvaluation} and \ref{tab:FundamentalStatistics} list the model accuracy evaluation and fundamental statistics of model parameters.

\subsection{\label{nonli} Nonlinearity measurement}
\begin{figure}[b]
\includegraphics[width=8.5cm]{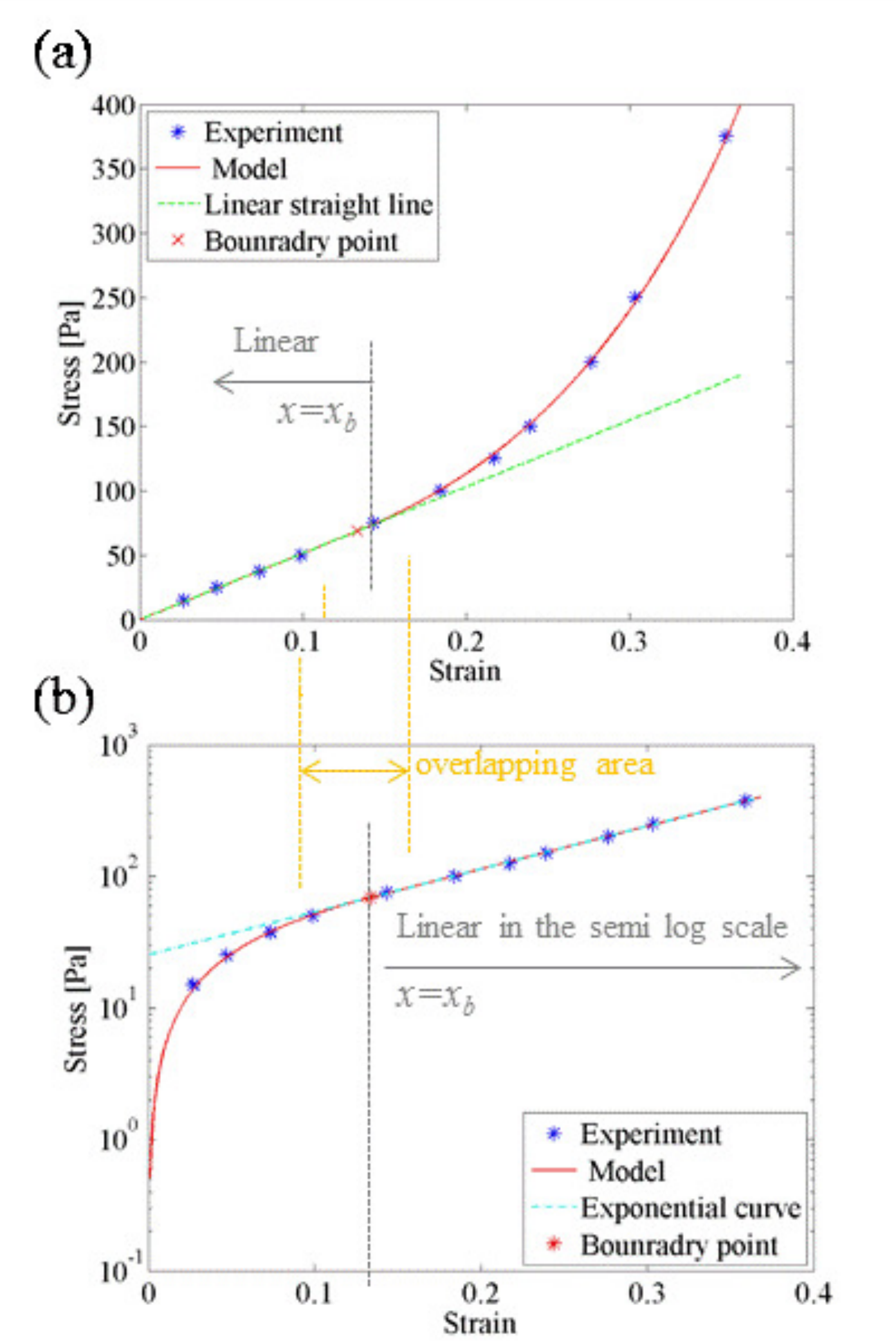}
\caption{\label{fig:NMF} (a) shows the relationship between strain \textit{$x_c$} and stress \textit{$f_c$} exhibit linear characteristics under low strain conditions. The stress nonlinearly increased during high strain conditions. The stress nonlinearly increased during high strain conditions.  (b) shows that the stress increase during high displacement is linear in the semi log scale; stress increases exponentially in the linear scale space. We found that linear straight line and nonlinear curves connected smoothly, with the exponential curve tangent to the linear straight region. The stress-strain relationship of our model, which fit the typical experimental results, are shown in red line. Our model and the experimental results are highly correlated. There is an overlapping area where both linearity and log scale linearity. Linearity holds to a certain degree over the boundary strain \textit{$x_b$}. log scale linearity holds to a certain degree before the boundary strain \textit{$x_b$}.} 
\end{figure}

Nonlinear properties of samples were investigated based on a series of creep tests under several applied stresses. Specifically, we looked at the relationship between the constant applied stress \textit{$f_c$} and the strain coefficient \textit{$x_c$} in a series of creep tests using several stresses. Typical experimental results for nonlinearity measurement of the sample are shown in Fig.\ref{fig:NMF}(a) and (b). Figure \ref{fig:NMF}(b) shows a semi-log diagram of the same data described in Fig.\ref{fig:NMF}(a). Figure \ref{fig:NMF}(a) shows the relationship between \textit{$x_c$} and \textit{$f_c$} exhibit linear characteristics under low strain conditions. The stress nonlinearly increased under high strain conditions. Figure \ref{fig:NMF}(b) shows that the stress increase during high displacement is linear in the semi log scale; stress increases exponentially in the linear scale space. We found that linear straight line and nonlinear curves connected smoothly, with the exponential curve tangent to the straight line region. 
We modeled nonlinear properties of the soft biological tissue based on these results and considerations, as shown in equation(\ref{eq:exponentialLinear}) and (\ref{eq:exponentialNonlinear}). Equation (\ref{eq:exponentialNonlinearLog}) is derived from log-log transformation of (\ref{eq:exponentialNonlinear}).

\begin{subequations}
\begin{eqnarray}
G{x_c} = {f_c}\quad \{ {x_c} < {x_b}\} 
\label{eq:exponentialLinear}. 
\\
{G_i} {e^{{G_n}{x_c}}} = {f_c}\quad \{ {x_c} > {x_b}\}
\label{eq:exponentialNonlinear}. 
\\
{G_n}{x_c} + \log {G_i} = \log {f_c}\quad \{ {x_c} > {x_b}\} 
\label{eq:exponentialNonlinearLog}. 
\end{eqnarray}
\end{subequations}

where \textit{$x_c$} is strain and \textit{$f_c$} is stress. \textit{G} is a proportionality factor (referred to as linear viscoelastic stiffness or linear stiffness herein), \textit{$x_b$} is the boundary strain between the linear and nonlinear range (called boundary strain), \textit{$G_n$} is a proportionality factor in the log space (called nonlinear viscoelastic stiffness or nonlinear stiffness), \textit{$G_i$} is the dependent parameter. Each parameter should fulfill the following relationship due to the exponential curve (\ref{eq:exponentialNonlinear}) being a tangent to the straight line (\ref{eq:exponentialLinear}). Details of this relationship are shown in Appendix \ref{ParameterDependency}.

\begin{subequations}
\begin{eqnarray}
{x_b} = \frac{1}{{{G_n}}}
\label{eq:parameter2} 
\\
{G_i} = \frac{G}{{{G_n}e}}
\label{eq:parameter3}. 
\end{eqnarray}
\end{subequations}

The parameters (\textit{$\alpha$}, \textit{G}, \textit{$G_n$}, \textit{$G_i$} and \textit{$x_b$}) of equations (\ref{eq:exponentialLinear}) and (\ref{eq:exponentialNonlinearLog}) were identified by fitting the experimental results for each sample. Thus, the equation (\ref{eq:exponentialLinear}) and (\ref{eq:exponentialNonlinearLog}) becomes (\ref{eq:parameter4}) and (\ref{eq:parameter5}) when the only independent parameters are used. 

\begin{subequations}
\begin{eqnarray}
G{x_c} = {f_c}\quad \{ {x_c} < \frac{1}{{{G_n}}}\}
\label{eq:parameter4}. 
\\
{G_n}{x_c} + \log (\frac{G}{{{G_n}e}}) = \log {f_c}\quad \{ {x_c} > \frac{1}{{{G_n}}}\}
\label{eq:parameter5}. 
\end{eqnarray}
\end{subequations}

We used the Extended Kalman Filter Algorithm to identify the parameters of equation (\ref{eq:parameter4}) and (\ref{eq:parameter5}) (ref. Appendix \ref{EKFforNonlinearityMeasurement} in detail), as  they are nonlinear simultaneous equations. The stress-strain relationship of our model, which fit the typical experimental results, are shown in Fig.\ref{fig:NMF}, show that our model and the experimental results are strongly correlated. Our model also fit all liver samples well. The coefficient of determination \textit{$R^2$} between our model and the experimental data in all samples was about 95\%. Tables \ref{tab:AccuracyEvaluation} and \ref{tab:FundamentalStatistics} list the model accuracy evaluation and fundamental statistics of model parameters.

Thus, we derived the nonlinear equations for our model shown in (\ref{eq:LinearFrac}) and (\ref{eq:NonlinearFrac}). We assume here that equations (\ref{eq:exponentialLinear}) and (\ref{eq:exponentialNonlinear}) can hold true more generally in the absence of creep tests.

\section{\label{discu}Discussion}
The main contribution of this article is the proposal of a Fractional Dynamics and Exponential Nonlinearity (FDEN) model to identify the rheological properties of soft biological tissue. We found from experimental results that biological tissues have specific properties: i) power law increases in storage elastic modulus and loss elastic modulus of the same slope; ii) power law gain and constant phase delay in frequency domain over two decades; iii) log-log scale linearity between time and strain relationships over two decades; and iv) linearity in low strain range and log scale linearity in high strain range between strain and stress relationships. The FDEN model uses only three dependent parameters (such as \textit{$\alpha$}, \textit{G} and \textit{$G_n$}) and represents the specific properties of soft biological tissues. The advantage of our model is that it strongly correlates with various experimental data, as shown in the section \ref{resul}. In addition, the small number of parameters used is valuable because it is suitable for parameter identification and inverse analysis. For example, the parameter identification methods in this article are basic, with only the Least Square Method (LSM) and Extended Kalman Filter (EKF) being used. Lastly, the meaning of each parameter is intuitively understood (for example, \textit{$\alpha$}: ratio of viscoelasticity, \textit{G}: linear stiffness, \textit{$G_n$}: nonlinear stiffness) and it is possible to compare the values with other tissue. The details of the discussion are described in the following sections.

\subsection{\label{ViscoelasticModelUsingFractionalCalculus} Viscoelastic model using fractional calculus}

We found from experimental results that soft biological tissues have specific properties, as described above in i)–-iv). Single terms in the fractional dynamics model (\ref{eq:LinearModelAgain}) represent the specific viscoelastic properties of soft biological tissues. 
Fractional calculus is an approach to mathematically describe natural phenomena that are related to viscoelastic behavior \cite{Baumann2011}. 

Fractional calculus is a branch of mathematical analysis concerned with taking real or complex number powers of differential operators. Fractional dynamics is a field of study in physics and mechanics concerned with investigating the behavior of objects and systems that are characterized by power law non-locality, power law long-term memory or fractal type properties by using integration and differentiation of non-integer orders, i.e., by fractional calculus methods \cite{Tarasov2013}. Fractional dynamics models have proven to be powerful tools in describing the dynamic behavior of various materials. The advantages of fractional dynamics models are their ability to describe real dynamic behavior and the fact they are simple enough for engineering calculations \cite{Pritz1996}. The equations for rheological models are generally based on stress-strain analyses and are traditionally represented with derivatives of integer order (ordinary differential equation). In other words, traditional methods to fit the viscoelastic response include several spring and dashpot elements. Recently, fractional dynamics models proved to be efficient in describing rheological materials such as rubber and tissues, reducing the number of parameters and showing a power law response \cite{Craiem2006}. 

Over the last years, fractional calculus has become an important tool in the analysis of viscoelastic materials composed of synthetic polymers \cite{Schiessel1995}. For example, Caputo et al. \cite{Caputo1971new,Caputo1971Linear,caputo1974vibrations} found good agreement with experimental results when using fractional calculation for the description of viscoelastic materials and established the connection between fractional calculation and the theory of linear viscoelasticity \cite{Schmidt2001}. Several authors \cite{Friedrich1991,Gloeckle1991} have also suggested the use of differential or integral equations of fractional order to describe viscoelastic behavior that is intermediate between purely elastic and purely viscous \cite{Heymans1994}. 

Although fractional calculus has wide application in describing the solid-liquid duality of synthetic polymers, it had until recently attracted limited attention in the field of biological materials, biomechanics, bio-rheology and cell viscoelasticity. Suki et al. \cite{Suki1994} found the pressure/volume response of a whole lung to be characterized by fractional calculus. Fractional calculus is also useful in bio-fields because many tissue-like materials (polymers, gels, emulsions, composites, and suspensions) exhibit power law responses to applied stress or strain \cite{Craiem2010}. Although such models are widely used for synthetic materials, they have been progressively substituted in the community by models relying on fractional calculus for biological materials \cite{Tanter2006}. Yuan et al. \cite{Klatt2010,Sack2009}, studied lung tissue and found its fractional order of evolution, while Chen et al. \cite{Yuan1997} applied the same model to agarose gels used for culturing tissues, particularly cartilage. An example of the power law behavior of elastic tissue was observed recently for viscoelastic measurements of blood vessels, where the analysis of these data was most conveniently performed using fractional order viscoelastic models \cite{Craiem2010}. Recently, the framework of fractional calculus has also been used in the research of magnetic resonance elastography \cite{Tanter2006,Klatt2010}. As above, fractional dynamics are gaining popularity in the field of viscoelasticity, with data and models already reported for the liver \cite{Kobayashi2005,Kobayashi2009,Kobayashi2012,kobayashi2012viscoelastic}, breast tissues \cite{Tsukune2014}, lung \cite{Suki1994,Yuan2000}, vessels \cite{Craiem2006}, muscle \cite{Kobayashi2012Soft,okamura2014study}, muscle cells \cite{Chen2004}, tendons \cite{Djordjevic2003} and blood cells \cite{Duenwald2009}. In short, research on fractional calculus has been applied widely to many fields, including biological materials.

The parameter \textit{$\alpha$} in fractional equation (\ref{eq:LinearModelAgain}) is in the order of a derivative that is commonly taken to range between 0 and 1. If \textit{$\alpha$} is 0, equation (\ref{eq:LinearModelAgain}) describes the behavior of a spring where \textit{G} specifies the springs’ stiffness. If \textit{$\alpha$} is 1, equation (\ref{eq:LinearModelAgain}) defines a dashpot, in which \textit{G} defines the viscosity. Thus, the fractional equation (\ref{eq:LinearModelAgain}) ‘interpolates’ between the material behavior of a spring and that of a dashpot \cite{Schmidt2001}. The rheological element that refers to equation (\ref{eq:LinearModelAgain}) was therefore introduced by Koeller  and termed a 'springpot' \cite{Craiem2010,Soczkiewicz2002}. As such, the derivative order \textit{$\alpha$} represents the index of viscosity of the system in the fractional dynamics model. A viscoelastic material is more governed by elastic properties than by the viscous properties when the derivative order \textit{$\alpha$} is close to 0. A viscoelastic material is more governed by viscous properties than by elastic properties when the derivative order \textit{$\alpha$} is close to 1.

 The value of the derivative order \textit{$\alpha$} was 0.125 (= 1/8) from the experimental results displayed in this article, indicating that the characteristics of soft biological tissue (liver) are intermediate between those of elastic and viscous bodies and is relatively close to an elastic material.

\subsection{\label{FractionalCalculusForTheDynamicViscoelasticAndCreepTests} Fractional calculus for the dynamic viscoelastic and creep tests}

In this article, the viscoelastic properties of soft biological tissues (liver) have been examined. The simple empirical equations describing strain creep (equation(\ref{eq:Creep}) and (\ref{eq:CreepLog})) have been put in a concise mathematical framework. We have chosen to describe viscoelasticity in terms of fractional calculation as in equation (\ref{eq:LinearModelAgain}). Certain important advantages must be emphasized, namely, fractional calculus has the following advantages \cite{Craiem2006,Suki1994}; i) fractional dynamics models accurately describe complex models with fewer parameters, ii) they improve curve fitting, principally with power law responses, iii) they allowed a physical justification in rheological and tissue samples. 

\subsubsection{\label{DynamicViscoelasticTestDiscussion} Dynamic viscoelastic test}
Measurements of mechanical complex impedance \textit{G*} in dynamic viscoelastic tests over a wide range of forcing frequencies $(10^{-1}$–-$10^{1}$ rad/s) in tissue sample revealed that the frequency dependence of rheological behavior represents a weak power law relationship over a wide range of frequencies ---For example, Fabry \cite{Fabry2003} reported that a weak power law relationship held over a range of frequencies ($10^{-2}$-–$10^{3}$ Hz) in muscle cells---. The storage modulus \textit{G'} increases with increasing frequency according to a weak power law with a power law exponent of approximately 0.125. The loss modulus G'' also follows the power law with a power law exponent of approximately 0.125. Fractional calculus provides a natural framework for describing such weak power law relationships \cite{Djordjevic2003}.

In contrast, mechanical models using ordinary differential equations have long been used, and their qualitative behavior is not representative of the actual behavior of materials. The characteristics of the frequency dependencies could be similar; however, the slopes of the experimental results do not fit those of the theoretical curves \cite{Pritz1996}. The shortcomings of ordinary differential models can be recognized by comparing the frequency curves exhibited by a material with those predicted by the models. The weak power law behavior cannot be accounted for by standard viscoelastic models characterized by ordinary differential equations \cite{Djordjevic2003}; i) storage modulus \textit{G'} should remain constant at low frequencies, which would indicate elastic behavior in ordinary differential equations and ii) loss storage modulus \textit{G''} increases and approaches a power law exponent of 1 at high frequencies, indicating viscous behavior in an ordinary differential equation. 

\subsubsection{\label{CreepTestDiscussion} Creep test}
Recent studies also indicated that the time domain data of tissues are well represented by a simple empirical equation involving a power law in time. Some studies also reported that creep responses represent power law stress to a step input in the time domain. Fung \cite{fung1981biomechanics} demonstrated in his theory that a distribution of time constants proportional to power of time over a finite range of time constants is appropriate for many tissues \cite{Suki1994}. Djordjevic and co-workers \cite{Djordjevic2003} reported that a parallel combination of a fractional calculus (springpot) and a dashpot properly predict the measured values for a rheological model of cultured smooth muscle cells \cite{Craiem2010}. As above, fractional calculus provides a natural framework for describing such power laws in the time domain \cite{Djordjevic2003}. 

In contrast, mechanical models using ordinary differential equations lack consistency between their qualitative behavior and the real behavior of materials curves. Although the characteristics of time dependences could be similar, the slopes of experimental results do not fit those of the theoretical curves. The shortcomings of ordinary differential models can be recognized by comparing the time domain curves observed for a material with those predicted by the models \cite{Pritz1996}. The power law behavior cannot be accounted for by standard viscoelastic models characterized using ordinary differential equations, such that; i) strain should remain constant at sufficient elapsed times, which would indicate elastic behavior in ordinary differential equations and ii) exponential increases in transient state, which would indicate viscous behavior in ordinary differential equations.
 
 \subsection{\label{ FractaslStructureAndTheFractionalLadderModel}  Fractal structure and the Fractional ladder model}

\begin{figure}[b]
\includegraphics[width=8.5cm]{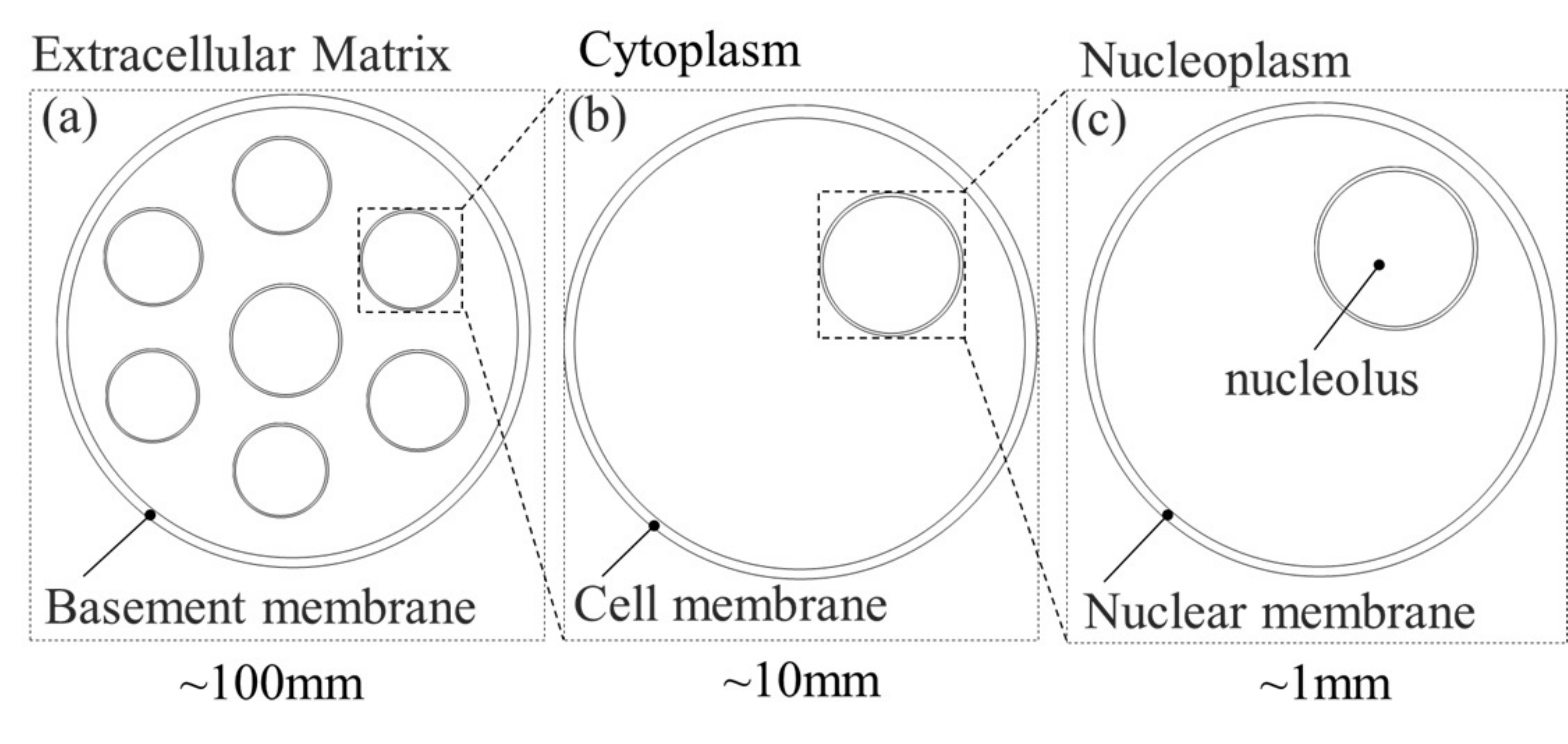}
\caption{\label{fig:Fractal} Fractal structure of soft biological tissues\cite{Kelly2009}. A repeating pattern with thin elastic membranes and viscous components is displayed at every scale. The first panel displays an infinite number of thin elastic membranes and viscous compartments. The second panel zooms in on the first panel, thus showing the self-similar layered structure. By allowing the number of structural components to extend indefinitely, the self-similarity of biological media is revealed.} 
\end{figure}

\begin{figure*}
\includegraphics[width=16cm]{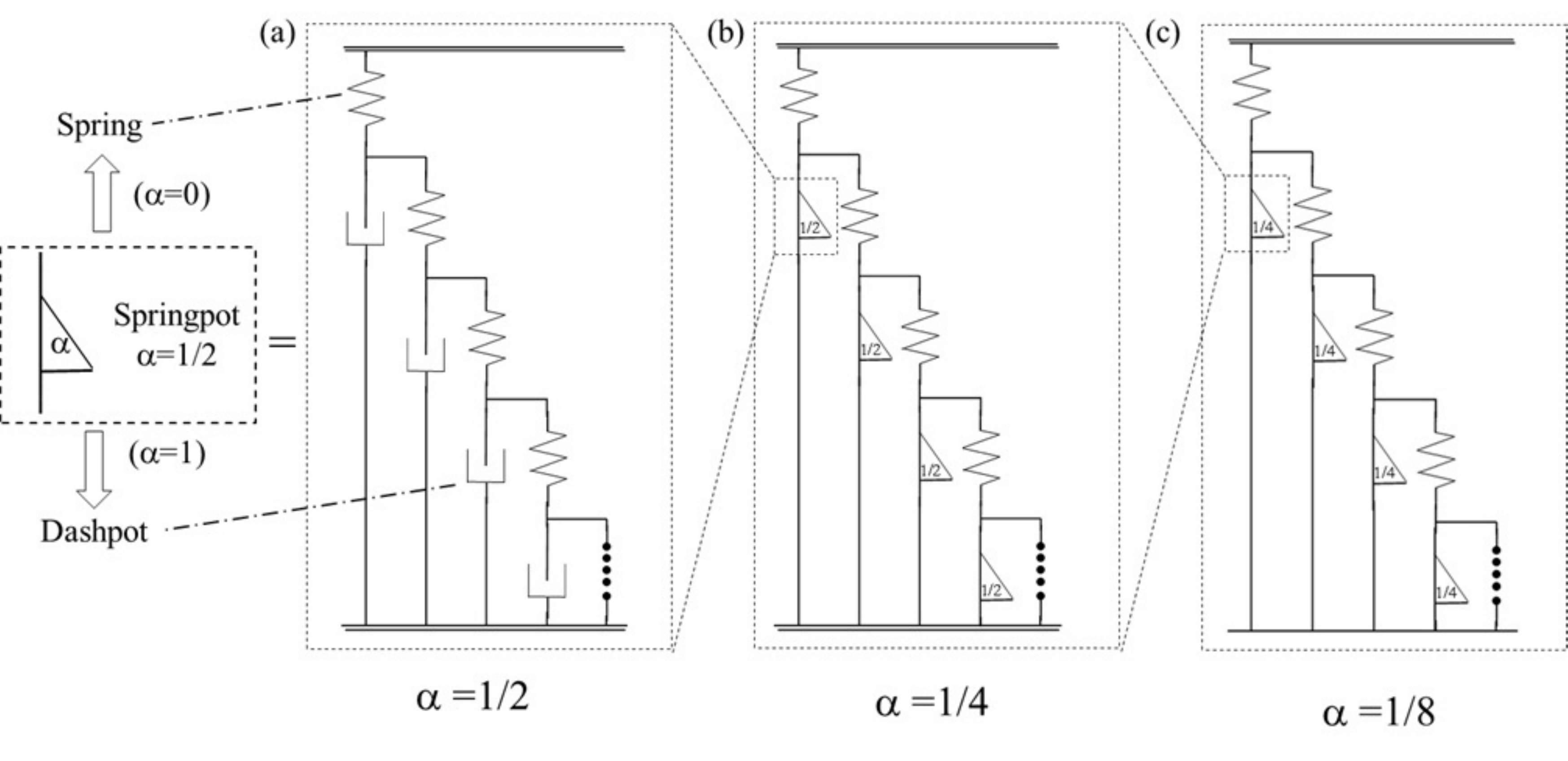}
\caption{\label{fig:LadderModel} Fractal ladder of springs and dashpots in order to capture these fractal components with the elastic membranes and viscous saline of biological tissue. Fractal tree networks were considered to model fractional calculus (orders of 1/2, 1/4 and 1/8 are described in (a), (b) and (c), respectively). These recursive ladder expansions provide various parameters of derivative order. Specifically, the ladder model can also be considered as a fundamental mechanical component of fractional derivative term, allowing more complicated fractal networks, or recursive ladders, to be constructed. The parameter \textit{$\alpha$} (order of derivative and also power law exponent) was about 1/8 (= 0.125) according to the experimental results of the dynamic viscoelastic tests in Figs.(\ref{fig:MI}) and (\ref{fig:BodeF}). This result suggests that liver tissue has a complex fractal structure such as in (c), where the fractional ladders are several times renormalized.
} 
\end{figure*}

Theoretical aspects of the fractional structure of soft biological tissue are partially explained with tissue fractal geometry in nature and its relationship to fractional calculus. Currently, fractal geometry and fractional calculus are applied to phenomenological theories for complex systems  \cite{Baumann2011}. Soft biological tissues also have fractal structures, such as in Fig.\ref{fig:Fractal}. A fractal is a natural phenomenon or a mathematical set that exhibits a repeating pattern displayed at every scale. For example, Schiessel and Blumen \cite{Schiessel1995}, Schiessel et al. \cite{Schiessel1995Fractal} and Heymans and Bauwens \cite{Heymans1994} have demonstrated that fractional equations, such as (\ref{eq:LinearModelAgain}), can be realized physically through the fractal structure of hierarchical arrangements of springs and dashpots like ladders \cite{Schiessel1999}. A related work \cite{Kelly2009} describes the details of the following layered fractal models of soft biological tissues based on the schema shown in Fig.\ref{fig:Fractal}. The first panel (a) displays an infinite number of thin elastic membranes and viscous compartments. The second panel (b) is a magnified region of the first panel (a) and the third panel (c) is a magnified region of the second panel(b), showing the self-similar layered structure. By allowing the number of structural components to extend indefinitely, the self-similarity of biological media is revealed. This topology is also depicted in Fig. \ref{fig:Fractal}, where the alternating elastic and viscous components are visualized as a self-similar hexagonal packing of spheres within spheres. In order to capture these fractal components with the elastic membranes and viscous saline of biological tissue, a fractal ladder of springs and dashpots is described in Fig.\ref{fig:LadderModel} \cite{Kelly2009}. A paper \cite{Kelly2009} described the properties of Fig. \ref{fig:LadderModel}(a), which presents the fractional derivative term with the order of derivative 1/2. Similar fractal tree networks were considered \cite{Kelly2009} to model the other order of fractional calculus (orders of 1/4 and 1/8 are also described in Fig. \ref{fig:LadderModel}(b) and (c), respectively). These recursive ladder expansions provide various parameters of derivative order. Specifically, the ladder model can also be considered as a fundamental mechanical component of fractional derivative term, allowing more complex fractal networks, or recursive ladders, to be constructed. For instance, consider a recursive ladder model constructed by replacing the viscous damper in Fig. \ref{fig:LadderModel}(a) with a fractal ladder, producing the arrangement shown in Fig. \ref{fig:LadderModel}(b) with the order of derivative 1/4. Similarly, a recursive ladder may be constructed by replacing the springs in Fig.\ref{fig:LadderModel}(b) with a fractal ladder, producing the arrangement shown in Fig.\ref{fig:LadderModel}(c) with the order of derivative 1/8.

The parameter  \textit{$\alpha$} (order of derivative and also power law exponent) was about 1/8 (= 0.125) according to the experimental results of the dynamic viscoelastic tests in Figs.(\ref{fig:MI}) and (\ref{fig:BodeF}). This suggests that liver tissue has a complex fractal structure, such as in Fig.\ref{fig:LadderModel}(c), where the fractional ladders are several times renormalized.

\subsection{\label{ MainContribution}  Main contribution}
From the viewpoint of fractional calculus, the main contribution of this study is to propose integration with fractional calculus and nonlinear equations, in other words, utilizing the nonlinearity of fractional calculus to describe soft biological tissue. The history of fractional dynamics in viscoelastic materials is long standing; however, the nonlinearity of the fractional equation for soft biological tissue has not yet been considered in related studies. The theoretical contribution of the present study is to empirically reveal the order of fractal structure of soft biological tissues (Fig.\ref{fig:LadderModel}(c)) via the analysis of the derivative order (\textit{$\alpha$}= 1/8). 

We believe that the fractional model with a single term is most suitable for the identification of bio-rheological properties, while many previous studies also have proposed serial and/or parallel arrangements of ordinary order models and fractional order models (such as a fractional generalized Voight model with a fractional term). The single fractional derivative term has the strong advantage of high model accuracy --– although the frequency range was relatively low in this study--– and the power law relationship is suitable for parameter identification, as described in the following section \ref{Time and frequency scale invariant}, \ref{Strain scale invariance} and \ref{Identification algorithm}.

\subsection{\label{ Exponential nonlinearity}  Exponential nonlinearity}
Figure \ref{fig:NMF} and equation (\ref{eq:exponentialNonlinear}) show that stress on soft biological tissue increases exponentially with strain. These trends are generally known in fields such as economics and in natural evolutionary processes. For example, value grows exponentially with time, technology has advanced at an exponential rate \cite{kurzweil2004law} (exponential growth of computing power is known as Moore’s law), market price in inflation shows exponential growth \cite{mizuno2002mechanism} and population growth (such as Malthusian Theory of Population) is exponential. The experimental results of stress and exponential models imply that the behaviors in the stress-strain relationship may have a similar mathematical structure, although the variable is not time, but, strain. In this theory, the exponential growth evolves due to a linear positive feedback mechanism, such as equation (\ref{eq:partial2}); an upward change in stress induces further stress increases rather than just incremental additons.

\begin{subequations}
\begin{eqnarray}
\left\{ {\begin{array}{*{20}{c}}
{\frac{{{d^\alpha }\varphi }}{{d{t^\alpha }}} = f}\\
{\frac{{{\partial ^2}\varphi }}{{\partial {x^2}}} = 0}
\end{array}} \right.\quad \{  - {x_b} < x < {x_b}\} 
\label{eq:partial1}. 
\\
\left\{ {\begin{array}{*{20}{c}}
{\frac{{{d^\alpha }\varphi }}{{d{t^\alpha }}} = f}\\
{\frac{{{\partial ^2}\varphi }}{{\partial {x^2}}} = G_n^2\varphi }
\end{array}} \right.\quad \{ x > {x_b}\}  \wedge \{ x <  - {x_b}\}
\label{eq:partial2}. 
\end{eqnarray}
\end{subequations}

\begin{figure}[b]
\includegraphics[width=8.5cm]{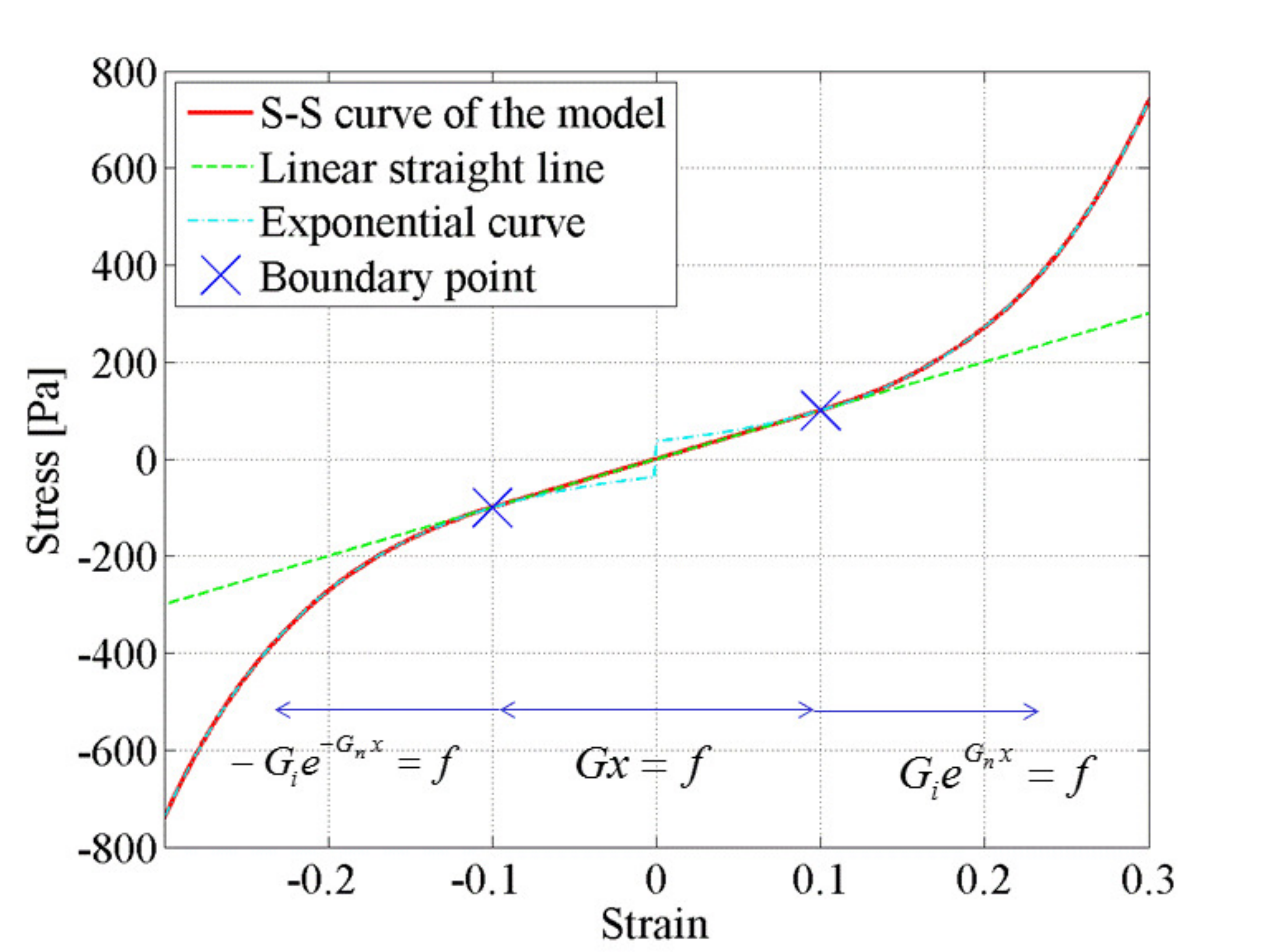}
\caption{\label{fig:PlusMinusNM} Negative and positive symmetry property of strain and stress relationship to introduce second order partial differential equation(in the case of \textit{$G=1000$} ad \textit{$G_n=10$}). The solution of first order partial differential equation only represent positive or negative exponential stress changes. The solution of second order partial differential equation represent positive and negative exponential stress changes such as this figure.
} 
\end{figure}

Where \textit{$\varphi$} is the intermediate variable between the upper and lower equations, which is related to force. We introduce a second-order partial differential equation here because of the negative and positive symmetry properties of the strain and stress relationship, as shown in Figure \ref{fig:PlusMinusNM} and equation (\ref{eq:partial7}). The solution for the second-order order partial differential equation is as follows; 


\begin{subequations}
\begin{eqnarray}
- {G_i}{e^{ - {G_n}x}} = f\quad\{ x< -{x_b}\}
\label{eq:partial4}. 
\\
Gx = f\quad \{  - {x_b} < x < {x_b}\}
\label{eq:partial5}. 
\\
{G_i}{e^{{G_{n{\kern 1pt} }}x}} = f\quad \{ x > {x_b}\}
\label{eq:partial6}. 
\end{eqnarray}
\label{eq:partial7}. 
\end{subequations}

The solution for the first-order partial differential equation only represents positive or negative exponential stress changes.

The smoothness of a fundamental mathematical function largely affects the robustness of identification, inverse analysis, structure analysis in computer simulations, and optimization using structure analysis. In particular, the smoothness at a point where mathematical function changes (specifically, \textit{x = $x_b$}) is important. Differentiability class is generally used in the classification of functions according to smoothness, more specifically, the properties of their derivatives. The differentiability class of our nonlinear model - between equations such as (\ref{eq:exponentialLinear}) and (\ref{eq:exponentialNonlinear}), (\ref{eq:partial4}) and (\ref{eq:partial5}), (\ref{eq:partial5}) and (\ref{eq:partial6})- is \textit{C1} at \textit{x} = \textit{$x_b$},\textit{$-x_b$}; this means curves are continuous and differentiable at \textit{x} = \textit{$x_b$}. In other words, curves are joined and first derivatives are continuous at \textit{x} = \textit{$x_b$}, \textit{$-x_b$}. Thus, the smoothness of function in our nonlinear model is maintained when compared with linear models. 

These smooth characteristics are an advantage of our model because the robustness of calculations in the boundary between linear and nonlinear characteristics is high. We can estimate the origin of strain from data in the nonlinear range due to the constraint condition of parameters described in equation (\ref{eq:parameter2}). Equation (\ref{eq:parameter2}) refers to subtangent (\textit{$x_b$}) as a word in geometry is constant in exponential function. The subtangent can be estimated using the tangent (\textit{$G_n$}) of the semi-log graph of the stress-strain relationship. This is an important characteristic because the zero strain point of soft biological tissue is generally difficult to define. The zero strain point is difficult to define because: 1) the zero strain point cannot be defined from linear data; 2) deformation of soft biological tissue is relatively large and is markedly affected by gravity force; and 3) viscoelasticity of soft biological tissue results in difficulties measuring the zero strain point.

\subsection{\label{Time and frequency scale invariant}  Time and frequency scale invariant}

One attribute of power laws is their scale invariance. Scaling the argument by a constant factor causes only a proportionate scaling of the function itself. Scaling by a constant simply multiplies the original power law relationship by the constant (parameter \textit{$\alpha$} in this article). Thus, it follows that all power laws with a particular scaling exponent are equivalent up to constant factors, as each is simply a scaled version of the others. There is no internal time scale that could typify the dynamics and no time characteristics are evident. 

Time scale invariance during creep tests can be described based on the above discussion, and creep response is not tied to any time scale; thus, it may be regarded as being scale-free. More specifically, as explained in this article, obtaining numerous time series data from creep tests is not necessary due to invariance in the time scale property. Only two data points at any time point are sufficient to robustly identify the parameter of equation (\ref{eq:LinearModelAgain}). Naturally, this is only a theory pertaining to identical conditions, and many data points are preferable to enhance the robustness of measurements. 

Frequency scale invariance during dynamic viscoelastic tests can be described in the same manner. There are no internal frequency scales that could typify the dynamics and no characteristic frequency was evident. Mechanical impedance responses are not tied to any frequency scale; thus, they may be regarded as being scale-free. More specifically, obtaining numerous frequency series data from dynamic viscoelastic tests are not necessary due to invariance in frequency scales. Only two data points at any frequency point are sufficient to robustly identify the parameter of equation (\ref{eq:LinearModelAgain}). Naturally, this is only a theory pertaining to identical conditions, and many data points are preferable to enhance the robustness of measurements.

\subsection{\label{Strain scale invariance}  Strain scale invariance}
Strain scale invariance –--while nonlinearity is not power law–-- also holds at the linear area and nonlinear scale. The relationship between stress and strain in the linear range exhibits strain scale invariance because of linearity. The relationship between the logarithms of stress and strain in the nonlinear range also exhibits strain scale invariance because of the linearity of the semi-log space. This scale invariance produces a strong relationship for identifying parameters. Theoretically speaking, a two point data set in the linear range (including stress = 0 and strain =0) is sufficient to identify linear stiffness \textit{G} (slope of stress and strain in the linear space). In addition, a two-point data set in the nonlinear range is sufficient to identify nonlinear stiffness \textit{$G_n$} (slope of stress and strain in the semi-log space). Of course, the aforementioned data set numbers are theoretical for identical situations, many data points are preferable to enhance measurement robustness.

 In addition, strict classifications between linear and nonlinear range should not be neccesary because of the smooth connectivity between the linear and nonlinear properties of soft biological tissues (Fig. {\ref{fig:NMF}). There is an overlapping area where both linearity and log scale linearity are held (about strain from 0.10 to 0.15 in Fig.{\ref{fig:NMF}). Linearity holds to a certain degree over the boundary strain \textit{$x_b$}. Furthermore, log scale linearity holds to a certain degree before the boundary strain \textit{$x_b$}. In other words, the data set in the overlapped area includes information from both the linear and nonlinear range. These properties, which are common to soft biological tissues and the FDEN model, are useful for identifying parameters and it is possible to identify both linear and nonlinear stiffness using only data sets derived from the overlapping area. However, further research is required to confirm this hypothesis.

\subsection{\label{Identification algorithm}  Identification algorithm}
Accoding to the simple model equation, its power law properties and its semi-log scale linearity, parameter identification in the FDEN model is simpler when compared to other reported models. The small number of parameters in the FDEN model contributes to a simple algorithm and parameter identification. These characteristics may be also effective in inverse analysis of computer structural simulations of tissue/organ deformation. 
For example, the parameter identification methods in this article are basic, with only the Least Square Method (LSM) and Extended Kalman Filter (EKF) being used. The parameters of the model in the creep test can be identified using LSM. EKF was used to identify the parameter of dynamic viscoelastic tests –--mechanical impedance--– because equations (\ref{eq:logGp}) and (\ref{eq:logGpp}) are nonlinear simultaneous equations. EKF was also used to identify the parameter of nonlinear models between stress and strain relationship because equations (\ref{eq:parameter4}) and (\ref{eq:parameter5}) are nonlinear simultaneous equations. 

The parameters of the model in the Bode diagram were identified via parameter identification of mechanical impedance in this article. It should be noted that LSM is sufficient to identify the value of \textit{$\alpha$} and \textit{G} in (\ref{eq:GainLog}). The correlation between model (\ref{eq:Phase}) and experimental data of the phase should be carefully checked in the case, as phase data can affect identification of the parameters.

\subsection{\label{Parameter variation}  Parameter variation}
Tables \ref{tab:AccuracyEvaluation} and \ref{tab:FundamentalStatistics} list the model accuracy evaluation and fundamental statistics of model parameters. Statistical analysis of the samples used in this study revealed that the maximum values of \textit{$\alpha$} and \textit{G} and \textit{$G_n$} were approximately 1, 4 and 3 times the minimum values, respectively. These results indicate that the linear stiffness \textit{G} and the nonlinear stiffness \textit{$G_n$} (also, \textit{$x_b$} as an independent parameter) has a large degree of variation when compared with ratio of viscoelasticity \textit{$\alpha$}. The one of limitation of this study is that the number of samples was insufficient to stastically analyze the variations in each parameter. We plan to study other tissue types in order to compare and discriminate between tissues using these static and variable parameters (for example, \cite{Tsukune2014}).

\begin{table*}
\caption{\label{tab:AccuracyEvaluation}
Accuracy evaluation results for the present model.}
\begin{ruledtabular}
\begin{tabular}{cccccccc}
\shortstack{Equation and \\ experimental data}&
\shortstack{Sample number\\(Trial number)}&
Avg. of \textit{$R^2$}&
Max. of \textit{$R^2$}&
Min. of \textit{$R^2$}&
S.D. of \textit{$R^2$}\\
\hline
Equation (\ref{eq:logGp})(\ref{eq:logGpp}) and dynamic viscoelastic test & 11 & 0.840 & 0.902 & 0.751 & 0.008 \\
Equation (\ref{eq:CreepLog}) and creep results & 64 (712) & 0.997 & 1.000 &0.950 &0.0073\\
Equation(\ref{eq:exponentialLinear})(\ref{eq:exponentialNonlinear}) and nonlinear measurement & 64 & 0.986 & 1.000 &0.923 & 0.018 \\
\end{tabular}
\end{ruledtabular}
\end{table*}

\begin{table*}
\caption{\label{tab:FundamentalStatistics}
Fundamental statistics of the parameters.}
\begin{ruledtabular}
\begin{tabular}{cccccccc}
Test&
Parameter&
\shortstack{Sample\\ number}&
Avg.&
Max.&
Min.&
S.D.\\
\hline
Dynamic viscoelastic test & \textit{G} & 11 & 391.1 & 518.2 & 248.2 & 111.2 \\
Dynamic viscoelastic test & \textit{$\alpha$} & 11 & 0.131 & 0.146 & 0.118 & 0.011 \\
Nonlinearity measurement & \textit{G}& 64 & 544.8 & 1294 &341.8&155.3\\
Nonlinearity measurement & \textit{$G_n$}& 64 & 8.547 & 13.18 &5.26 &1.604\\
Nonlinearity measurement & \textit{$x_b$}&64 & 0.121 & 0.190 &0.076 &0.0022\\
Nonlinearity measurement & \textit{$G_i$}&64 & 23.90 & 39.64 &11.35 &6.206\\
\end{tabular}
\end{ruledtabular}
\end{table*}

\subsection{\label{Limitations} Limitations}
The main limitation of this study is that we only measured and evaluated liver samples. Similar evaluations must be performed with other tissues in order to clarify the universality of the FDEN model. This will allow us to clarify the applicability of our model to various tissue types. We believe that the FDEN model can represent other biological tissues consisting of a single tissue type, excluding non-soft tissues such as bone and tissue exhibiting plasticity, such as brain. Our previous nonlinear viscoelastic model with four parameters  \cite{Kobayashi2005,Kobayashi2009,Kobayashi2012,kobayashi2012viscoelastic} has already been partially evaluated with breast tissues (fibro-glandur, fat, muscle) \cite{Kobayashi2012Enhanced,Tsukune2014}. We plan to evaluate the FDEN model using other tissues in future studies. 

In addition, this article did not utilize stress relaxation and indentation tests, which are basic experiments to evaluate viscoelastic properties and stress-strain nonlinearity, respectively. Parameter identification in the FDEN model using these tests are described in appendices \ref{RelaxationTest} and \ref{IndentationTest}. Stress relaxation analysis using fractional viscoelasticity was introduced in related studies  \cite{Friedrich1991,Gloeckle1991}, as well as in our work on human stretch applications \cite{okamura2014study}. Moreover, indentation (in the case of needle insertion and palapation for medical robotics) using the nonlinear model has been introduced in related studies\cite{Kobayashi2005,Kobayashi2009,Kobayashi2012,kobayashi2012viscoelastic,Kobayashi2012Enhanced,Tsukune2014}. 

\section{\label{Conclusion} Conclusion}
We proposed a simple empirical model using Fractional Dynamics and Exponential Nonlinearity (FDEN) to identify the rheological properties of soft biological tissue. The model is derived from detailed material measurements using samples isolated from porcine liver. We conducted dynamic viscoelastic tests and creep tests on liver samples using a rheometer. The experimental results indicated that biological tissue has specific properties, such as: i) power law increases in storage elastic modulus and loss elastic modulus with the same slope; ii) power law gain decrease and constant phase delay in the frequency domain over two decades; iii) log-log scale linearity between time and strain relationships under constant force; and iv) linear and log scale linearity between strain and stress relationships. Our simple FDEN model uses only three dependent parameters and represents the specific properties of soft biological tissue.

\begin{acknowledgments}
This work was supported in part by JST PRESTO, Japan; in part by the Global Centers of Excellence (GCOE) Program and Grants for Excellent Graduate Schools “Global Robot Academia” of Waseda University, Tokyo, Japan; and in part by a Grant-in-Aid of Scientific Research from MEXT (No. 25350577), Japan;.
\end{acknowledgments}
\appendix

\section{\label{ShearStressAndStrain}Calculation of shear stress and strain}
Torque \textit{T} applied to the sample, and the torsional angle \textit{$\theta$} of the sample, were measured using a rheometer (AR-G2 or AR550; TA Instruments, New Castle, DE). From these measurements, the conventional shear strain \textit{x} and conventional shear stress \textit{f} were calculated using Eq. (\ref{eq:appA1a}) and (\ref{eq:appA1b}), respectively:

\begin{subequations}
\begin{eqnarray}
f =  \frac{1}{2}\ \frac{2}{{\pi {R^3}}}T
\label{eq:appA1a} 
\\
x = \frac{1}{2}\ \frac{R}{d}\theta
\label{eq:appA1b} 
\end{eqnarray}
\end{subequations}

where \textit{d} and \textit{R} are the length and radius of the cylinder (ref. Fig.\ref{fig:rheometer}), respectively. The mean stress and strain on the sample (half values of outer stress and strain on the sample) are referred to in the experimental results, because they are adequate for consideration of the nonlinear properties. \textit{R} was 20 [mm] and d was 5 [mm] in the experimental setup of this paper.

\section{\label{ParameterDependency}Parameter dependency}
We modeled nonlinear properties of soft biological tissue based on these results and considerations, as shown in Eq. (\ref{eq:appB1a}) and (\ref{eq:appB1b}). 

\begin{subequations}
\begin{eqnarray}
Gx = f\quad \{ x < {x_b}\} 
\label{eq:appB1a} 
\\
{G_i} {e^{{G_n}x}} = f\quad \{ x > {x_b}\} 
\label{eq:appB1b} 
\end{eqnarray}
\end{subequations}

where \textit{x} is strain and \textit{f} is stress. \textit{G} is linear stiffness, \textit{$x_b$} is the boundary strain, \textit{$G_n$} is nonlinear stiffness, \textit{$G_i$} is the dependent parameter, and \textit{e} is Napier's constant. Each parameter should fulfill the condition that the exponential curve (\ref{eq:appB1b} is a tangent to the straight line (\ref{eq:appB1a}) at \textit{$x=x_b$}.
(\ref{eq:appB2a}) and (\ref{eq:appB2b}) is derived by derivation of equation (\ref{eq:appB1a}) and (\ref{eq:appB1b});

\begin{subequations}
\begin{eqnarray}
\frac{{df}}{{dx}} = G\quad \{ x < {x_b}\} 
\label{eq:appB2a} 
\\
\frac{{df}}{{dx}} = {G_n}({G_i}{\kern 1pt} {e^{{G_n}{\kern 1pt} x}})\quad \{ x > {x_b}\} 
\label{eq:appB2b} 
\end{eqnarray}
\end{subequations}

The following equation may thus be fulfilled by plugging \textit{$x=x_b$} into (\ref{eq:appB1a}) and (\ref{eq:appB1b}).

\begin{eqnarray}
 {G_i}{e^{{G_n}{x_b}}} = G{x_b}
\label{eq:appB3} 
\end{eqnarray}

Moreover, the following equation may be fulfilled by plugging \textit{$x=x_b$} into (\ref{eq:appB2a}) and (\ref{eq:appB2b}).

\begin{eqnarray}
 {G_n}({G_i} {e^{{G_n}{x_b}}} )= G
\label{eq:appB4} 
\end{eqnarray}

By plugging the left side of equation (\ref{eq:appB3}) into (\ref{eq:appB4}),

\begin{eqnarray}
{G_n}{x_b} = 1
\label{eq:appB5} 
\end{eqnarray}

Then, 

\begin{eqnarray}
{x_b} = \frac{1}{{{G_n}}}
\label{eq:appB6} 
\end{eqnarray}

By plugging (\ref{eq:appB6}) into (\ref{eq:appB4});

\begin{eqnarray}
{G_n}{G_i}e = G
\label{eq:appB7} 
\end{eqnarray}

Then, 

\begin{eqnarray}
{G_i} = \frac{G}{{{G_n}e}}
\label{eq:appB8} 
\end{eqnarray}

Each parameter may then fulfill the above relationship, particularly (\ref{eq:appB6}) and ((\ref{eq:appB8}). The stress-strain relationship with several \textit{$G_n$} in our model is described in Fig.\ref{fig:ParameterConstrain} to represent the meaning of parameter constrain between \textit{$G_n$} and \textit{$x_b$}. 

\begin{figure}[b]
\includegraphics[width=8.5cm]{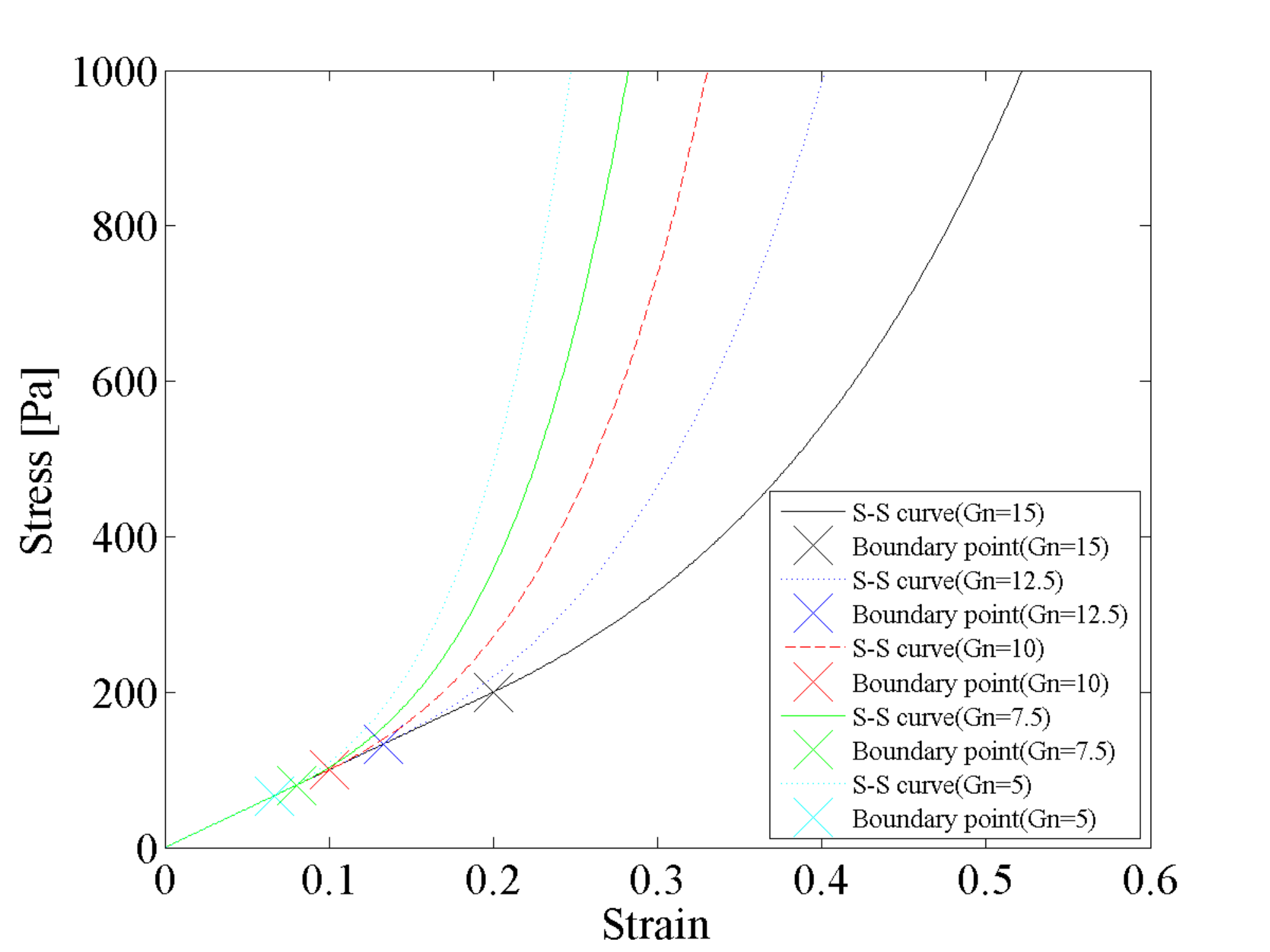}
\caption{\label{fig:ParameterConstrain} The stress-strain relationship with several \textit{$G_n$} in our model to provide the meaning of parameter constrain between \textit{$G_n$} and \textit{$x_b$}. \textit{G} was set to 1000 Pa.} 
\end{figure}

\section{\label{EKFforDynamicViscoelasticTest}Extended Kalman Filter (EKF) for Dynamic viscoelastic test}

This section shows the methodology used to identify the paramerter described in section \ref{mecha}. The model for dynamic viscoelastic test was as follows from equations (\ref{eq:logGp})-(\ref{eq:logGpp});

\begin{subequations}
\begin{eqnarray}
\log G' = \alpha \log \omega + \log G{'_0}
\label{eq:AppC1a}. 
\\
\log G'' = \alpha \log \omega  + \log G'{'_0}
\label{eq:AppC1b} 
\end{eqnarray}
\end{subequations}

where \textit{G'}, \textit{G''} and \textit{$\omega$} are variables; \textit{$\alpha$} and \textit{G} are parameters. 

We obtained the set of \textit{G'} and \textit{G''} at each value for frequency \textit{$\omega$} from the experiment. We identified the parameter from these data using EKF for system identification. The system identification in EKF is generally described as follows; 

\begin{subequations}
\begin{eqnarray}
{\theta _{k + 1}} = f({\theta _k},{\psi _k})
\label{eq:AppC2a} 
\\
{y_k} = g({\theta _k},{\zeta _k})
\label{eq:AppC2b} 
\end{eqnarray}
\end{subequations}

where \textit{k = 0, 1, 2,…} represents the discrete iteration index (number of data set in this case), \textit{$\theta$} is the n-dimensional state vector, \textit{$\psi$} is the n-dimensional system noise vector, \textit{y }is the p-dimensional observation vector, \textit{$\zeta$} is the p-dimensional observation noise vector, and \textit{f(·)} and \textit{g(·)} are the nonlinear vector functions. In the theory of state-space, (\ref{eq:AppC2a}) and (\ref{eq:AppC2b}) are known as the system model (or state model) and the observation model, respectively. 

Parameter vector is regarded as a state vector in EKF for system identification. Where the state vector (parameter vector) \textit{$\theta$} is a constant vector and the observation noise vector \textit{$\zeta$} is a Gaussian white noise with zero mean, (\ref{eq:AppC2a}) and (\ref{eq:AppC2b}) are represented as:

\begin{subequations}
\begin{eqnarray}
{\theta _{k + 1}} = I{\theta _k}
\label{eq:AppC3a} 
\\
{y_k} = h({\theta _k}) + {\zeta _k}
\label{eq:AppC3b} 
\end{eqnarray}
\end{subequations}

where \textit{I} is the identity matrix and \textit{h(·)} is the nonlinear vector function. In the case of system identification for dynamic viscoelastic test, the state vector (parameter vector), \textit{$\theta$} observation vector \textit{y} and the nonlinear vector function \textit{h(·)} are particulary regarded as follows; 

\begin{subequations}
\begin{eqnarray}
\theta  &=& \left[ {\begin{array}{*{20}{c}}\alpha \\G\end{array}} \right]
\label{eq:AppC4a} 
\\
y &=& \left[ {\begin{array}{*{20}{c}}
{\log G'}\\
{\log G''}
\end{array}} \right]
\label{eq:AppC4b} 
\\
h(\theta ) &=& \left[ {\begin{array}{*{20}{c}}
{\alpha \log \omega + \log (G\cos (\frac{\pi }{2}\alpha ))}\\
{\alpha \log \omega + \log (G\sin (\frac{\pi }{2}\alpha ))}
\end{array}} \right]
\label{eq:AppC4c} 
\end{eqnarray}
\end{subequations}

The EKF algorithm (ref.\cite{Hoshi2008}) using (\ref{eq:AppC4a})-(\ref{eq:AppC4c}) was applied to identify the parameter from the data set. It was not necessary to set initial values for each parameter \textit{$\theta_0$}, meaning that \textit{$\theta_0$} was zero vector. 

\section{\label{EKFforNonlinearityMeasurement}Extended Kalman Filter (EKF) for Nonlinearity measurement}

This section shows the methodology used to identify the paramerter described in section \ref{nonli}. The model for nonlinearity measurment is as follows from equations (\ref{eq:parameter4})-(\ref{eq:parameter5});

\begin{subequations}
\begin{eqnarray}
G{x_c} &=& {f_c}\quad \{ {x_c} < \frac{1}{{{G_n}}}\}
\label{eq:AppD1a}. 
\\
{G_n}{x_c} + \log (\frac{G}{{{G_n}e}}) &=& \log ({f_c})\quad \{ {x_c} > \frac{1}{{{G_n}}}\}
\label{eq:AppD1b}. 
\end{eqnarray}
\end{subequations}

where \textit{$f_c$} and \textit{$x_c$} are variables, \textit{G} and \textit{$G_n$} are parameters, \textit{e} is Neipia’s contant. 
We obtained the set of \textit{$f_c$} at each strain \textit{$x_c$} from the experiment. We identified the parameter from these data using EKF for system identificatrion. The algorithm to identify the parameter is the same as in Appendix \ref{EKFforDynamicViscoelasticTest}, particularly equations (\ref{eq:AppC2a})-(\ref{eq:AppC3b}). In the case of system identification for nonlinearity measurment, the state vector (parameter vector) \textit{$\theta$}, observation vector \textit{y} and nonlinear vector function \textit{h()} are regarded as follows;

\begin{subequations}
\begin{eqnarray}
\theta  &=& \left[ {\begin{array}{*{20}{c}}
G\\
{{G_n}}
\end{array}} \right]
\label{eq:AppD2a}. 
\\
y &=& \left[ {\begin{array}{*{20}{c}}
f&{\{ x < \frac{1}{{{G_n}}}\} }\\
{\log f}&{\{ x > \frac{1}{{{G_n}}}\} }
\end{array}} \right]
\label{eq:AppD2b}. 
\\
h(\theta ) &=& \left[ {\begin{array}{*{20}{c}}
{Gx}&{\{ x < \frac{1}{{{G_n}}}\} }\\
{{G_n}x + \log (\frac{G}{{{G_n}e}})}&{\{ x > \frac{1}{{{G_n}}}\} }
\end{array}} \right]
\label{eq:AppD2c}. 
\end{eqnarray}
\end{subequations}

The EKF algorithm (ref. \cite{Hoshi2008}) using (\ref{eq:AppD2a})-(\ref{eq:AppD2c}) was applied to identify the parameter from the data set. Each data set affected a single term of the vector, where the upper term for the vectors was updated via low strain data, and the lower term for the vectors was updated via high strain data. Initial values for each parameter \textit{$\theta_0$} needed to be explicitly set in the case of nonlinearity measurement. Therefore, we first approximated the parameters to set initial values of each parameter \textit{$\theta_0$}. We used only low strain data (three data set from minimum strain) for approximation of the parameter \textit{G}, while we used only high strain data (three data set from maximum strain) for approximation of parameter \textit{$G_n$}. These approximations of the parameters can be identified using LSM –--simple linear regression of equations (\ref{eq:AppD1a}) and (\ref{eq:AppD1b})---.

\section{\label{RelaxationTest}Relaxation test}

A model equation of stress in relaxation tests can be devised as follows. We assumed that equation (\ref{eq:LinearModelAgain}) is valid for a relaxation test, while nonlinearity was evaluated by a series of relaxation tests under several applied strains. Specifically, equation (\ref{eq:LinearModelAgain}) becomes (\ref{eq:AppE1a}) if (\ref{eq:LinearModelAgain}) is solved for the conditions of the relaxation test. Here, the applied strain is constant \textit{$x_c$}. Equation (\ref{eq:AppE1b}) is derived from the log-log transformation of (\ref{eq:AppE1a}).

\begin{subequations}
\begin{eqnarray}
f = G\Gamma (1 + r){x_c}{t^{ - \alpha }}( = {f_c}{t^{ - \alpha }})
\label{eq:AppE1a}. 
\\
\log f =  - \alpha \log t + \log {f_c}
\label{eq:AppE1b}. 
\end{eqnarray}
\end{subequations}

where \textit{f} is stress and \textit{t} is time; \textit{$x_c$} is constant strain; \textit{G} is linear stiffness at each strain, and \textit{$\Gamma()$} is the gamma function. \textit{$f_c$} is the coefficient determining the strain value as a parameter, which is defined as follows:

\begin{eqnarray}
{f_c} = G\Gamma (1 + \alpha ){x_c}
\label{eq:AppE2}. 
\end{eqnarray}

The LSM algorithm can be used to identify the parameters of equation (\ref{eq:AppE1b}) –-linear regression–- for each sample. We calculated the other independent parameter \textit{G} via equation (\ref{eq:AppE2}).
Nonlinear properties of samples can be investigated based on a series of relaxation tests under several applied strains. Specifically, nonlinearity measures the relationship between the constant applied strain \textit{$x_c$} and the stress coefficient \textit{$f_c$} in the series of relaxation tests under several strains. 

\section{\label{IndentationTest}Indentation test}
A model equation for indentation test –--the reaction force measurement during constant velocity strain change--– can be theoretically calculated as follows. It should be noted that the sensitivity of parameter \textit{$\alpha$} from the steady-state of reaction force \textit{f} is generally low in experiments with soft biological tissue. Linear stiffness \textit{G} and nonlinear stiffness \textit{$G_n$} can thus be identified from reaction force on indentation test, when the value of parameter \textit{$\alpha$} is roughly known. 
We assumed that equations (\ref{eq:LinearFrac}) and (\ref{eq:NonlinearFrac}) –--the FDEN model--– are valid for indentation test, when the value of parameter \textit{$\alpha$} is already known. We collected time seies data for force \textit{ f(t)} on the conditions of indentation test. Here, the applied strain is \textit{x(t)}= \textit{$v_ot$}. Equations (\ref{eq:LinearFrac}) - (\ref{eq:NonlinearFrac}) become identical to static problems such as (\ref{eq:exponentialLinear})-(\ref{eq:exponentialNonlinear}) and (\ref{eq:AppF1a})-(\ref{eq:AppF1a}) when the fractional integrated stress \textit{f'} is considered as follows:

\begin{subequations}
\begin{eqnarray}
Gx = f'= {D^{( - \alpha )}}f\quad \{ x < {x_b}\} 
\label{eq:AppF1a}. 
\\
{G_n}{x} + \log {G_i} = \log {f'}={D^{( - \alpha )}}log{f} \{{x_c}>{x_b}\}
\label{eq:AppF1b}. 
\end{eqnarray}
\end{subequations}

where D(\textit{$\alpha$}) refers to \textit{$\alpha$}th-order derivative and D(-\textit{$\alpha$}) refers to \textit{$\alpha$}th-order integral.
The numerical fractional integration, which is necessary in the above calculation is introduced in various studies (for example, \cite{Ma2004}). Parameters (\textit{G}, \textit{$G_n$}, \textit{$G_i$} and \textit{$x_b$}) of equations (\ref{eq:AppF1a}) and (\ref{eq:AppF1b}) can be identified via the same method introduced in Section 3.C, while we use fractional integrated stress \textit{f'} on behalf of stress \textit{f}.

\nocite{*}

\bibliographystyle{apsrev}
\bibliography{Manuscript}

\begin{thebibliography}{61}
\expandafter\ifx\csname natexlab\endcsname\relax\def\natexlab#1{#1}\fi
\expandafter\ifx\csname bibnamefont\endcsname\relax
  \def\bibnamefont#1{#1}\fi
\expandafter\ifx\csname bibfnamefont\endcsname\relax
  \def\bibfnamefont#1{#1}\fi
\expandafter\ifx\csname citenamefont\endcsname\relax
  \def\citenamefont#1{#1}\fi
\expandafter\ifx\csname url\endcsname\relax
  \def\url#1{\texttt{#1}}\fi
\expandafter\ifx\csname urlprefix\endcsname\relax\def\urlprefix{URL }\fi
\providecommand{\bibinfo}[2]{#2}
\providecommand{\eprint}[2][]{\url{#2}}

\bibitem[{\citenamefont{Fung}(1981)}]{fung1981biomechanics}
\bibinfo{author}{\bibfnamefont{Y.}~\bibnamefont{Fung}},
  \emph{\bibinfo{title}{Biomechanics: mechanical properties of living
  tissues}}, Biomechanics / Y. C. Fung (\bibinfo{publisher}{Springer-Verlag},
  \bibinfo{year}{1981}), ISBN \bibinfo{isbn}{9780387904726},
  \urlprefix\url{http://books.google.de/books?id=HkhRAAAAMAAJ}.

\bibitem[{\citenamefont{Miller and Chinzei}(2002)}]{Miller2002}
\bibinfo{author}{\bibfnamefont{K.}~\bibnamefont{Miller}} \bibnamefont{and}
  \bibinfo{author}{\bibfnamefont{K.}~\bibnamefont{Chinzei}},
  \bibinfo{journal}{Journal of Biomechanics} \textbf{\bibinfo{volume}{35}},
  \bibinfo{pages}{483} (\bibinfo{year}{2002}), ISSN \bibinfo{issn}{0021-9290}.

\bibitem[{\citenamefont{Sarver et~al.}(2003)\citenamefont{Sarver, Robinson, and
  Elliott}}]{Sarver2003}
\bibinfo{author}{\bibfnamefont{J.~J.} \bibnamefont{Sarver}},
  \bibinfo{author}{\bibfnamefont{P.~S.} \bibnamefont{Robinson}},
  \bibnamefont{and} \bibinfo{author}{\bibfnamefont{D.~M.}
  \bibnamefont{Elliott}}, \bibinfo{journal}{Journal of biomechanical
  engineering} \textbf{\bibinfo{volume}{125}}, \bibinfo{pages}{754}
  (\bibinfo{year}{2003}), ISSN \bibinfo{issn}{01480731}.

\bibitem[{\citenamefont{Brands et~al.}(2004)\citenamefont{Brands, Peters, and
  Bovendeerd}}]{Brands2004}
\bibinfo{author}{\bibfnamefont{D.~W.~a.} \bibnamefont{Brands}},
  \bibinfo{author}{\bibfnamefont{G.~W.~M.} \bibnamefont{Peters}},
  \bibnamefont{and} \bibinfo{author}{\bibfnamefont{P.~H.~M.}
  \bibnamefont{Bovendeerd}}, \bibinfo{journal}{Journal of Biomechanics}
  \textbf{\bibinfo{volume}{37}}, \bibinfo{pages}{127} (\bibinfo{year}{2004}),
  ISSN \bibinfo{issn}{00219290}.

\bibitem[{\citenamefont{Chui et~al.}(2004)\citenamefont{Chui, Kobayashi, Chen,
  Hisada, and Sakuma}}]{chui2004combined}
\bibinfo{author}{\bibfnamefont{C.}~\bibnamefont{Chui}},
  \bibinfo{author}{\bibfnamefont{E.}~\bibnamefont{Kobayashi}},
  \bibinfo{author}{\bibfnamefont{X.}~\bibnamefont{Chen}},
  \bibinfo{author}{\bibfnamefont{T.}~\bibnamefont{Hisada}}, \bibnamefont{and}
  \bibinfo{author}{\bibfnamefont{I.}~\bibnamefont{Sakuma}},
  \bibinfo{journal}{Medical and Biological Engineering and Computing}
  \textbf{\bibinfo{volume}{42}}, \bibinfo{pages}{787} (\bibinfo{year}{2004}).

\bibitem[{\citenamefont{Kerdok et~al.}(2006)\citenamefont{Kerdok, Ottensmeyer,
  and Howe}}]{kerdok2006effects}
\bibinfo{author}{\bibfnamefont{A.~E.} \bibnamefont{Kerdok}},
  \bibinfo{author}{\bibfnamefont{M.~P.} \bibnamefont{Ottensmeyer}},
  \bibnamefont{and} \bibinfo{author}{\bibfnamefont{R.~D.} \bibnamefont{Howe}},
  \bibinfo{journal}{Journal of Biomechanics} \textbf{\bibinfo{volume}{39}},
  \bibinfo{pages}{2221} (\bibinfo{year}{2006}).

\bibitem[{\citenamefont{Sedef et~al.}(2006)\citenamefont{Sedef, Samur, and
  Basdogan}}]{Sedef}
\bibinfo{author}{\bibfnamefont{M.}~\bibnamefont{Sedef}},
  \bibinfo{author}{\bibfnamefont{E.}~\bibnamefont{Samur}}, \bibnamefont{and}
  \bibinfo{author}{\bibfnamefont{C.}~\bibnamefont{Basdogan}},
  \bibinfo{journal}{2006 14th Symposium on Haptic Interfaces for Virtual
  Environment and Teleoperator Systems} pp. \bibinfo{pages}{201--208}
  (\bibinfo{year}{2006}).

\bibitem[{\citenamefont{Samur et~al.}(2007)\citenamefont{Samur, Sedef,
  Basdogan, Avtan, and Duzgun}}]{Samur2007}
\bibinfo{author}{\bibfnamefont{E.}~\bibnamefont{Samur}},
  \bibinfo{author}{\bibfnamefont{M.}~\bibnamefont{Sedef}},
  \bibinfo{author}{\bibfnamefont{C.}~\bibnamefont{Basdogan}},
  \bibinfo{author}{\bibfnamefont{L.}~\bibnamefont{Avtan}}, \bibnamefont{and}
  \bibinfo{author}{\bibfnamefont{O.}~\bibnamefont{Duzgun}},
  \bibinfo{journal}{Medical Image Analysis} \textbf{\bibinfo{volume}{11}},
  \bibinfo{pages}{361} (\bibinfo{year}{2007}), ISSN \bibinfo{issn}{13618415}.

\bibitem[{\citenamefont{Tanaka et~al.}(2008)\citenamefont{Tanaka, Higashimori,
  and Kaneko}}]{Tanaka2008}
\bibinfo{author}{\bibfnamefont{N.}~\bibnamefont{Tanaka}},
  \bibinfo{author}{\bibfnamefont{M.}~\bibnamefont{Higashimori}},
  \bibnamefont{and} \bibinfo{author}{\bibfnamefont{M.}~\bibnamefont{Kaneko}},
  \bibinfo{journal}{Engineering in Medicine and Biology Society, Annual
  International Conference of the IEEE} \textbf{\bibinfo{volume}{2008}},
  \bibinfo{pages}{106} (\bibinfo{year}{2008}), ISSN \bibinfo{issn}{1557-170X}.

\bibitem[{\citenamefont{Gao et~al.}(2010)\citenamefont{Gao, Lister, and
  Desai}}]{gao2010constitutive}
\bibinfo{author}{\bibfnamefont{Z.}~\bibnamefont{Gao}},
  \bibinfo{author}{\bibfnamefont{K.}~\bibnamefont{Lister}}, \bibnamefont{and}
  \bibinfo{author}{\bibfnamefont{J.~P.} \bibnamefont{Desai}},
  \bibinfo{journal}{Annals of biomedical engineering}
  \textbf{\bibinfo{volume}{38}}, \bibinfo{pages}{505} (\bibinfo{year}{2010}).

\bibitem[{\citenamefont{Sadeghi~Naini et~al.}(2011)\citenamefont{Sadeghi~Naini,
  Patel, and Samani}}]{SadeghiNaini2011}
\bibinfo{author}{\bibfnamefont{A.}~\bibnamefont{Sadeghi~Naini}},
  \bibinfo{author}{\bibfnamefont{R.~V.} \bibnamefont{Patel}}, \bibnamefont{and}
  \bibinfo{author}{\bibfnamefont{A.}~\bibnamefont{Samani}},
  \bibinfo{journal}{IEEE Transactions on Biomedical Engineering}
  \textbf{\bibinfo{volume}{58}}, \bibinfo{pages}{2852} (\bibinfo{year}{2011}),
  ISSN \bibinfo{issn}{00189294}.

\bibitem[{\citenamefont{Darvish and Crandall}(2001)}]{Darvish2001}
\bibinfo{author}{\bibfnamefont{K.~K.} \bibnamefont{Darvish}} \bibnamefont{and}
  \bibinfo{author}{\bibfnamefont{J.~R.} \bibnamefont{Crandall}},
  \bibinfo{journal}{Medical engineering \& physics}
  \textbf{\bibinfo{volume}{23}}, \bibinfo{pages}{633} (\bibinfo{year}{2001}),
  ISSN \bibinfo{issn}{1350-4533}.

\bibitem[{\citenamefont{Kim et~al.}(2003)\citenamefont{Kim, Tay, Stylopoulos,
  Rattner, and Srinivasan}}]{kim2003characterization}
\bibinfo{author}{\bibfnamefont{J.}~\bibnamefont{Kim}},
  \bibinfo{author}{\bibfnamefont{B.~K.} \bibnamefont{Tay}},
  \bibinfo{author}{\bibfnamefont{N.}~\bibnamefont{Stylopoulos}},
  \bibinfo{author}{\bibfnamefont{D.~W.} \bibnamefont{Rattner}},
  \bibnamefont{and} \bibinfo{author}{\bibfnamefont{M.~A.}
  \bibnamefont{Srinivasan}}, in \emph{\bibinfo{booktitle}{Medical Image
  Computing and Computer Assisted Intervention 2003}}
  (\bibinfo{publisher}{Springer}, \bibinfo{year}{2003}), pp.
  \bibinfo{pages}{206--213}.

\bibitem[{\citenamefont{Schwartz et~al.}(2005)\citenamefont{Schwartz,
  Denninger, Rancourt, Moisan, and Laurendeau}}]{schwartz2005modelling}
\bibinfo{author}{\bibfnamefont{J.-M.} \bibnamefont{Schwartz}},
  \bibinfo{author}{\bibfnamefont{M.}~\bibnamefont{Denninger}},
  \bibinfo{author}{\bibfnamefont{D.}~\bibnamefont{Rancourt}},
  \bibinfo{author}{\bibfnamefont{C.}~\bibnamefont{Moisan}}, \bibnamefont{and}
  \bibinfo{author}{\bibfnamefont{D.}~\bibnamefont{Laurendeau}},
  \bibinfo{journal}{Medical Image Analysis} \textbf{\bibinfo{volume}{9}},
  \bibinfo{pages}{103} (\bibinfo{year}{2005}).

\bibitem[{\citenamefont{Kim and Srinivasan}(2005)}]{Kim2005}
\bibinfo{author}{\bibfnamefont{J.}~\bibnamefont{Kim}} \bibnamefont{and}
  \bibinfo{author}{\bibfnamefont{M.~a.} \bibnamefont{Srinivasan}},
  \bibinfo{journal}{Medical image computing and computer-assisted intervention}
  \textbf{\bibinfo{volume}{8}}, \bibinfo{pages}{599} (\bibinfo{year}{2005}),
  ISSN \bibinfo{issn}{3540293264}.

\bibitem[{\citenamefont{Liu et~al.}(2007)\citenamefont{Liu, Noonan, Zweiri,
  Althoefer, Seneviratne et~al.}}]{liu2007development}
\bibinfo{author}{\bibfnamefont{H.}~\bibnamefont{Liu}},
  \bibinfo{author}{\bibfnamefont{D.~P.} \bibnamefont{Noonan}},
  \bibinfo{author}{\bibfnamefont{Y.~H.} \bibnamefont{Zweiri}},
  \bibinfo{author}{\bibfnamefont{K.}~\bibnamefont{Althoefer}},
  \bibinfo{author}{\bibfnamefont{L.~D.} \bibnamefont{Seneviratne}},
  \bibnamefont{et~al.}, in \emph{\bibinfo{booktitle}{Intelligent Robots and
  Systems, 2007. IROS 2007. IEEE/RSJ International Conference on}}
  (\bibinfo{organization}{IEEE}, \bibinfo{year}{2007}), pp.
  \bibinfo{pages}{208--213}.

\bibitem[{\citenamefont{Famaey and Sloten}(2008)}]{famaey2008soft}
\bibinfo{author}{\bibfnamefont{N.}~\bibnamefont{Famaey}} \bibnamefont{and}
  \bibinfo{author}{\bibfnamefont{J.~V.} \bibnamefont{Sloten}},
  \bibinfo{journal}{Computer methods in biomechanics and biomedical
  engineering} \textbf{\bibinfo{volume}{11}}, \bibinfo{pages}{351}
  (\bibinfo{year}{2008}).

\bibitem[{\citenamefont{Lu et~al.}(2010)\citenamefont{Lu, Zhu, Richmond, and
  Middleton}}]{Lu2010}
\bibinfo{author}{\bibfnamefont{Y.~T.} \bibnamefont{Lu}},
  \bibinfo{author}{\bibfnamefont{H.~X.} \bibnamefont{Zhu}},
  \bibinfo{author}{\bibfnamefont{S.}~\bibnamefont{Richmond}}, \bibnamefont{and}
  \bibinfo{author}{\bibfnamefont{J.}~\bibnamefont{Middleton}},
  \bibinfo{journal}{Journal of Biomechanics} \textbf{\bibinfo{volume}{43}},
  \bibinfo{pages}{2629} (\bibinfo{year}{2010}), ISSN \bibinfo{issn}{00219290}.

\bibitem[{\citenamefont{Sims et~al.}(2010)\citenamefont{Sims, Stait-Gardner,
  Fong, Morley, Price, Hoffman, Simmons, and Schindhelm}}]{Sims2010}
\bibinfo{author}{\bibfnamefont{a.~M.} \bibnamefont{Sims}},
  \bibinfo{author}{\bibfnamefont{T.}~\bibnamefont{Stait-Gardner}},
  \bibinfo{author}{\bibfnamefont{L.}~\bibnamefont{Fong}},
  \bibinfo{author}{\bibfnamefont{J.~W.} \bibnamefont{Morley}},
  \bibinfo{author}{\bibfnamefont{W.~S.} \bibnamefont{Price}},
  \bibinfo{author}{\bibfnamefont{M.}~\bibnamefont{Hoffman}},
  \bibinfo{author}{\bibfnamefont{a.}~\bibnamefont{Simmons}}, \bibnamefont{and}
  \bibinfo{author}{\bibfnamefont{K.}~\bibnamefont{Schindhelm}},
  \bibinfo{journal}{Biomechanics and Modeling in Mechanobiology}
  \textbf{\bibinfo{volume}{9}}, \bibinfo{pages}{703} (\bibinfo{year}{2010}),
  ISSN \bibinfo{issn}{16177959}.

\bibitem[{\citenamefont{Marchesseau et~al.}(2010)\citenamefont{Marchesseau,
  Heimann, Chatelin, Willinger, and Delingette}}]{Marchesseau2010}
\bibinfo{author}{\bibfnamefont{S.}~\bibnamefont{Marchesseau}},
  \bibinfo{author}{\bibfnamefont{T.}~\bibnamefont{Heimann}},
  \bibinfo{author}{\bibfnamefont{S.}~\bibnamefont{Chatelin}},
  \bibinfo{author}{\bibfnamefont{R.}~\bibnamefont{Willinger}},
  \bibnamefont{and}
  \bibinfo{author}{\bibfnamefont{H.}~\bibnamefont{Delingette}},
  \bibinfo{journal}{Progress in Biophysics and Molecular Biology}
  \textbf{\bibinfo{volume}{103}}, \bibinfo{pages}{185} (\bibinfo{year}{2010}),
  ISSN \bibinfo{issn}{00796107}.

\bibitem[{\citenamefont{Ahn and Kim}(2010)}]{Ahn2010}
\bibinfo{author}{\bibfnamefont{B.}~\bibnamefont{Ahn}} \bibnamefont{and}
  \bibinfo{author}{\bibfnamefont{J.}~\bibnamefont{Kim}},
  \bibinfo{journal}{Medical Image Analysis} \textbf{\bibinfo{volume}{14}},
  \bibinfo{pages}{138} (\bibinfo{year}{2010}), ISSN \bibinfo{issn}{13618415}.

\bibitem[{\citenamefont{Basafa and Farahmand}(2011)}]{Basafa2011}
\bibinfo{author}{\bibfnamefont{E.}~\bibnamefont{Basafa}} \bibnamefont{and}
  \bibinfo{author}{\bibfnamefont{F.}~\bibnamefont{Farahmand}},
  \bibinfo{journal}{International Journal of Computer Assisted Radiology and
  Surgery} \textbf{\bibinfo{volume}{6}}, \bibinfo{pages}{297}
  (\bibinfo{year}{2011}), ISSN \bibinfo{issn}{18616410}.

\bibitem[{\citenamefont{Kobayashi et~al.}(2005)\citenamefont{Kobayashi,
  Okamoto, and Fujie}}]{Kobayashi2005}
\bibinfo{author}{\bibfnamefont{Y.~K.~Y.} \bibnamefont{Kobayashi}},
  \bibinfo{author}{\bibfnamefont{J.~O.~J.} \bibnamefont{Okamoto}},
  \bibnamefont{and} \bibinfo{author}{\bibfnamefont{M.}~\bibnamefont{Fujie}},
  \bibinfo{journal}{Proceedings of the 2005 IEEE International Conference on
  Robotics and Automation} pp. \bibinfo{pages}{1644--1651}
  (\bibinfo{year}{2005}), ISSN \bibinfo{issn}{10504729}.

\bibitem[{\citenamefont{Kobayashi et~al.}(2009)\citenamefont{Kobayashi, Onishi,
  Hoshi, Kawamura, Hashizume, and Fujie}}]{Kobayashi2009}
\bibinfo{author}{\bibfnamefont{Y.}~\bibnamefont{Kobayashi}},
  \bibinfo{author}{\bibfnamefont{A.}~\bibnamefont{Onishi}},
  \bibinfo{author}{\bibfnamefont{T.}~\bibnamefont{Hoshi}},
  \bibinfo{author}{\bibfnamefont{K.}~\bibnamefont{Kawamura}},
  \bibinfo{author}{\bibfnamefont{M.}~\bibnamefont{Hashizume}},
  \bibnamefont{and} \bibinfo{author}{\bibfnamefont{M.~G.} \bibnamefont{Fujie}},
  \bibinfo{journal}{International Journal of Computer Assisted Radiology and
  Surgery} \textbf{\bibinfo{volume}{4}}, \bibinfo{pages}{53}
  (\bibinfo{year}{2009}), ISSN \bibinfo{issn}{18616410}.

\bibitem[{\citenamefont{Kobayashi
  et~al.}(2012{\natexlab{a}})\citenamefont{Kobayashi, Kato, Watanabe, Hoshi,
  Kawamura, and Fujie}}]{Kobayashi2012}
\bibinfo{author}{\bibfnamefont{Y.}~\bibnamefont{Kobayashi}},
  \bibinfo{author}{\bibfnamefont{A.}~\bibnamefont{Kato}},
  \bibinfo{author}{\bibfnamefont{H.}~\bibnamefont{Watanabe}},
  \bibinfo{author}{\bibfnamefont{T.}~\bibnamefont{Hoshi}},
  \bibinfo{author}{\bibfnamefont{K.}~\bibnamefont{Kawamura}}, \bibnamefont{and}
  \bibinfo{author}{\bibfnamefont{M.~G.} \bibnamefont{Fujie}},
  \bibinfo{journal}{Journal of Biomechanical Science and Engineering}
  \textbf{\bibinfo{volume}{7}}, \bibinfo{pages}{177}
  (\bibinfo{year}{2012}{\natexlab{a}}), ISSN \bibinfo{issn}{18809863}.

\bibitem[{\citenamefont{Kobayashi
  et~al.}(2012{\natexlab{b}})\citenamefont{Kobayashi, Watanabe, Hoshi,
  Kawamura, and Fujie}}]{kobayashi2012viscoelastic}
\bibinfo{author}{\bibfnamefont{Y.}~\bibnamefont{Kobayashi}},
  \bibinfo{author}{\bibfnamefont{H.}~\bibnamefont{Watanabe}},
  \bibinfo{author}{\bibfnamefont{T.}~\bibnamefont{Hoshi}},
  \bibinfo{author}{\bibfnamefont{K.}~\bibnamefont{Kawamura}}, \bibnamefont{and}
  \bibinfo{author}{\bibfnamefont{M.~G.} \bibnamefont{Fujie}}, in
  \emph{\bibinfo{booktitle}{Soft Tissue Biomechanical Modeling for Computer
  Assisted Surgery}} (\bibinfo{publisher}{Springer},
  \bibinfo{year}{2012}{\natexlab{b}}), pp. \bibinfo{pages}{41--67}.

\bibitem[{\citenamefont{Kobayashi
  et~al.}(2012{\natexlab{c}})\citenamefont{Kobayashi, Suzuki, Kato, Hatano,
  Konishi, Hashizume, and Fujie}}]{Kobayashi2012Enhanced}
\bibinfo{author}{\bibfnamefont{Y.}~\bibnamefont{Kobayashi}},
  \bibinfo{author}{\bibfnamefont{M.}~\bibnamefont{Suzuki}},
  \bibinfo{author}{\bibfnamefont{A.}~\bibnamefont{Kato}},
  \bibinfo{author}{\bibfnamefont{M.}~\bibnamefont{Hatano}},
  \bibinfo{author}{\bibfnamefont{K.}~\bibnamefont{Konishi}},
  \bibinfo{author}{\bibfnamefont{M.}~\bibnamefont{Hashizume}},
  \bibnamefont{and} \bibinfo{author}{\bibfnamefont{M.~G.} \bibnamefont{Fujie}},
  \bibinfo{journal}{IEEE Transactions on Robotics}
  \textbf{\bibinfo{volume}{28}}, \bibinfo{pages}{710}
  (\bibinfo{year}{2012}{\natexlab{c}}), ISSN \bibinfo{issn}{15523098}.

\bibitem[{\citenamefont{Tsukune et~al.}(2014)\citenamefont{Tsukune, Kobayashi,
  Miyashita, and Fujie}}]{Tsukune2014}
\bibinfo{author}{\bibfnamefont{M.}~\bibnamefont{Tsukune}},
  \bibinfo{author}{\bibfnamefont{Y.}~\bibnamefont{Kobayashi}},
  \bibinfo{author}{\bibfnamefont{T.}~\bibnamefont{Miyashita}},
  \bibnamefont{and} \bibinfo{author}{\bibfnamefont{G.~M.} \bibnamefont{Fujie}},
  \bibinfo{journal}{International Journal of Computer Assisted Radiology and
  Surgery} \textbf{\bibinfo{volume}{10}}, \bibinfo{pages}{593}
  (\bibinfo{year}{2014}), ISSN \bibinfo{issn}{1861-6410}.

\bibitem[{\citenamefont{Kobayashi
  et~al.}(2012{\natexlab{d}})\citenamefont{Kobayashi, Watanabe, Seki, Ando, and
  Fujie}}]{Kobayashi2012Soft}
\bibinfo{author}{\bibfnamefont{Y.}~\bibnamefont{Kobayashi}},
  \bibinfo{author}{\bibfnamefont{T.}~\bibnamefont{Watanabe}},
  \bibinfo{author}{\bibfnamefont{M.}~\bibnamefont{Seki}},
  \bibinfo{author}{\bibfnamefont{T.}~\bibnamefont{Ando}}, \bibnamefont{and}
  \bibinfo{author}{\bibfnamefont{M.~G.} \bibnamefont{Fujie}},
  \bibinfo{journal}{Advanced Robotics} \textbf{\bibinfo{volume}{26}},
  \bibinfo{pages}{1253} (\bibinfo{year}{2012}{\natexlab{d}}), ISSN
  \bibinfo{issn}{0169-1864}.

\bibitem[{\citenamefont{Okamura et~al.}(2014)\citenamefont{Okamura, Tsukune,
  Kobayashi, and Fujie}}]{okamura2014study}
\bibinfo{author}{\bibfnamefont{N.}~\bibnamefont{Okamura}},
  \bibinfo{author}{\bibfnamefont{M.}~\bibnamefont{Tsukune}},
  \bibinfo{author}{\bibfnamefont{Y.}~\bibnamefont{Kobayashi}},
  \bibnamefont{and} \bibinfo{author}{\bibfnamefont{M.~G.} \bibnamefont{Fujie}},
  in \emph{\bibinfo{booktitle}{Engineering in Medicine and Biology Society,
  Annual International Conference of the IEEE}} (\bibinfo{organization}{IEEE},
  \bibinfo{year}{2014}), pp. \bibinfo{pages}{6919--6922}.

\bibitem[{\citenamefont{Baumann}(2011)}]{Baumann2011}
\bibinfo{author}{\bibfnamefont{G.}~\bibnamefont{Baumann}},
  \bibinfo{journal}{Fractals in Biology and Medicine} pp.
  \bibinfo{pages}{17--30--30} (\bibinfo{year}{2011}).

\bibitem[{\citenamefont{Tarasov}(2013)}]{Tarasov2013}
\bibinfo{author}{\bibfnamefont{V.~E.} \bibnamefont{Tarasov}},
  \bibinfo{journal}{International Journal of Modern Physics B}
  \textbf{\bibinfo{volume}{27}}, \bibinfo{pages}{1330005}
  (\bibinfo{year}{2013}), ISSN \bibinfo{issn}{0217-9792},
  \eprint{arXiv:1502.07681v1}.

\bibitem[{\citenamefont{Pritz}(1996)}]{Pritz1996}
\bibinfo{author}{\bibfnamefont{T.}~\bibnamefont{Pritz}},
  \bibinfo{journal}{Journal of Sound and Vibration}
  \textbf{\bibinfo{volume}{195}}, \bibinfo{pages}{103} (\bibinfo{year}{1996}),
  ISSN \bibinfo{issn}{0022460X}.

\bibitem[{\citenamefont{Craiem and Armentano}(2006)}]{Craiem2006}
\bibinfo{author}{\bibfnamefont{D.~O.} \bibnamefont{Craiem}} \bibnamefont{and}
  \bibinfo{author}{\bibfnamefont{R.~L.} \bibnamefont{Armentano}},
  \bibinfo{journal}{Engineering in Medicine and Biology Society, Annual
  International Conference of the IEEE} pp. \bibinfo{pages}{1098--1101}
  (\bibinfo{year}{2006}), ISSN \bibinfo{issn}{05891019}.

\bibitem[{\citenamefont{Schiessel and
  Blumen}(1995{\natexlab{a}})}]{Schiessel1995}
\bibinfo{author}{\bibfnamefont{H.}~\bibnamefont{Schiessel}} \bibnamefont{and}
  \bibinfo{author}{\bibfnamefont{a.}~\bibnamefont{Blumen}},
  \bibinfo{journal}{Macromolecules} \textbf{\bibinfo{volume}{28}},
  \bibinfo{pages}{4013} (\bibinfo{year}{1995}{\natexlab{a}}), ISSN
  \bibinfo{issn}{0024-9297}.

\bibitem[{\citenamefont{Caputo and
  Mainardi}(1971{\natexlab{a}})}]{Caputo1971new}
\bibinfo{author}{\bibfnamefont{M.}~\bibnamefont{Caputo}} \bibnamefont{and}
  \bibinfo{author}{\bibfnamefont{F.}~\bibnamefont{Mainardi}},
  \bibinfo{journal}{Pure and Applied Geophysics PAGEOPH}
  \textbf{\bibinfo{volume}{91}}, \bibinfo{pages}{134}
  (\bibinfo{year}{1971}{\natexlab{a}}).

\bibitem[{\citenamefont{Caputo and
  Mainardi}(1971{\natexlab{b}})}]{Caputo1971Linear}
\bibinfo{author}{\bibfnamefont{M.}~\bibnamefont{Caputo}} \bibnamefont{and}
  \bibinfo{author}{\bibfnamefont{F.}~\bibnamefont{Mainardi}},
  \bibinfo{journal}{La Rivista del Nuovo Cimento} \textbf{\bibinfo{volume}{1}},
  \bibinfo{pages}{161} (\bibinfo{year}{1971}{\natexlab{b}}), ISSN
  \bibinfo{issn}{0393697X}.

\bibitem[{\citenamefont{Caputo}(1974)}]{caputo1974vibrations}
\bibinfo{author}{\bibfnamefont{M.}~\bibnamefont{Caputo}}, \bibinfo{journal}{The
  Journal of the Acoustical Society of America} \textbf{\bibinfo{volume}{56}},
  \bibinfo{pages}{897} (\bibinfo{year}{1974}).

\bibitem[{\citenamefont{Schmidt and Gaul}(2001)}]{Schmidt2001}
\bibinfo{author}{\bibfnamefont{a.}~\bibnamefont{Schmidt}} \bibnamefont{and}
  \bibinfo{author}{\bibfnamefont{L.}~\bibnamefont{Gaul}},
  \bibinfo{journal}{Constitutive models for rubber}  (\bibinfo{year}{2001}).

\bibitem[{\citenamefont{Friedrich}(1991)}]{Friedrich1991}
\bibinfo{author}{\bibfnamefont{C.}~\bibnamefont{Friedrich}},
  \bibinfo{journal}{Rheologica Acta} \textbf{\bibinfo{volume}{30}},
  \bibinfo{pages}{151} (\bibinfo{year}{1991}), ISSN \bibinfo{issn}{00354511}.

\bibitem[{\citenamefont{Gloeckle and Nonnenmacher}(1991)}]{Gloeckle1991}
\bibinfo{author}{\bibfnamefont{W.~G.} \bibnamefont{Gloeckle}} \bibnamefont{and}
  \bibinfo{author}{\bibfnamefont{T.~F.} \bibnamefont{Nonnenmacher}},
  \bibinfo{journal}{Macromolecules} \textbf{\bibinfo{volume}{24}},
  \bibinfo{pages}{6426} (\bibinfo{year}{1991}), ISSN \bibinfo{issn}{0024-9297}.

\bibitem[{\citenamefont{Heymans and Bauwens}(1994)}]{Heymans1994}
\bibinfo{author}{\bibfnamefont{N.}~\bibnamefont{Heymans}} \bibnamefont{and}
  \bibinfo{author}{\bibfnamefont{J.~C.} \bibnamefont{Bauwens}},
  \bibinfo{journal}{Rheologica Acta} \textbf{\bibinfo{volume}{33}},
  \bibinfo{pages}{210} (\bibinfo{year}{1994}), ISSN \bibinfo{issn}{00354511}.

\bibitem[{\citenamefont{Suki et~al.}(1994)\citenamefont{Suki, Barab\'{a}si, and
  Lutchen}}]{Suki1994}
\bibinfo{author}{\bibfnamefont{B.}~\bibnamefont{Suki}},
  \bibinfo{author}{\bibfnamefont{a.~L.} \bibnamefont{Barab\'{a}si}},
  \bibnamefont{and} \bibinfo{author}{\bibfnamefont{K.~R.}
  \bibnamefont{Lutchen}}, \bibinfo{journal}{Journal of applied physiology
  (Bethesda, Md. : 1985)} \textbf{\bibinfo{volume}{76}}, \bibinfo{pages}{2749}
  (\bibinfo{year}{1994}), ISSN \bibinfo{issn}{8750-7587}.

\bibitem[{\citenamefont{Craiem and Magin}(2010)}]{Craiem2010}
\bibinfo{author}{\bibfnamefont{D.}~\bibnamefont{Craiem}} \bibnamefont{and}
  \bibinfo{author}{\bibfnamefont{R.~L.} \bibnamefont{Magin}},
  \bibinfo{journal}{Physical biology} \textbf{\bibinfo{volume}{7}},
  \bibinfo{pages}{13001} (\bibinfo{year}{2010}), ISSN
  \bibinfo{issn}{1478-3975}.

\bibitem[{\citenamefont{Tanter et~al.}(2006)\citenamefont{Tanter, Fink, Robert,
  Sinkus, and Larrat}}]{Tanter2006}
\bibinfo{author}{\bibfnamefont{M.}~\bibnamefont{Tanter}},
  \bibinfo{author}{\bibfnamefont{M.}~\bibnamefont{Fink}},
  \bibinfo{author}{\bibfnamefont{B.}~\bibnamefont{Robert}},
  \bibinfo{author}{\bibfnamefont{R.}~\bibnamefont{Sinkus}}, \bibnamefont{and}
  \bibinfo{author}{\bibfnamefont{B.}~\bibnamefont{Larrat}},
  \bibinfo{journal}{2006 IEEE Ultrasonics Symposium} pp.
  \bibinfo{pages}{1033--1036} (\bibinfo{year}{2006}), ISSN
  \bibinfo{issn}{1051-0117}.

\bibitem[{\citenamefont{Klatt et~al.}(2010)\citenamefont{Klatt, Papazoglou,
  Braun, and Sack}}]{Klatt2010}
\bibinfo{author}{\bibfnamefont{D.}~\bibnamefont{Klatt}},
  \bibinfo{author}{\bibfnamefont{S.}~\bibnamefont{Papazoglou}},
  \bibinfo{author}{\bibfnamefont{J.}~\bibnamefont{Braun}}, \bibnamefont{and}
  \bibinfo{author}{\bibfnamefont{I.}~\bibnamefont{Sack}},
  \bibinfo{journal}{Physics in medicine and biology}
  \textbf{\bibinfo{volume}{55}}, \bibinfo{pages}{6445} (\bibinfo{year}{2010}),
  ISSN \bibinfo{issn}{0031-9155}.

\bibitem[{\citenamefont{Sack et~al.}(2009)\citenamefont{Sack, Beierbach,
  Wuerfel, Klatt, Hamhaber, Papazoglou, Martus, and Braun}}]{Sack2009}
\bibinfo{author}{\bibfnamefont{I.}~\bibnamefont{Sack}},
  \bibinfo{author}{\bibfnamefont{B.}~\bibnamefont{Beierbach}},
  \bibinfo{author}{\bibfnamefont{J.}~\bibnamefont{Wuerfel}},
  \bibinfo{author}{\bibfnamefont{D.}~\bibnamefont{Klatt}},
  \bibinfo{author}{\bibfnamefont{U.}~\bibnamefont{Hamhaber}},
  \bibinfo{author}{\bibfnamefont{S.}~\bibnamefont{Papazoglou}},
  \bibinfo{author}{\bibfnamefont{P.}~\bibnamefont{Martus}}, \bibnamefont{and}
  \bibinfo{author}{\bibfnamefont{J.}~\bibnamefont{Braun}},
  \bibinfo{journal}{NeuroImage} \textbf{\bibinfo{volume}{46}},
  \bibinfo{pages}{652} (\bibinfo{year}{2009}), ISSN \bibinfo{issn}{10538119}.

\bibitem[{\citenamefont{Yuan et~al.}(1997)\citenamefont{Yuan, Ingenito, and
  Suki}}]{Yuan1997}
\bibinfo{author}{\bibfnamefont{H.}~\bibnamefont{Yuan}},
  \bibinfo{author}{\bibfnamefont{E.~P.} \bibnamefont{Ingenito}},
  \bibnamefont{and} \bibinfo{author}{\bibfnamefont{B.}~\bibnamefont{Suki}},
  \bibinfo{journal}{Journal of applied physiology (Bethesda, Md. : 1985)}
  \textbf{\bibinfo{volume}{83}}, \bibinfo{pages}{1420} (\bibinfo{year}{1997}),
  ISSN \bibinfo{issn}{8750-7587}.

\bibitem[{\citenamefont{Yuan et~al.}(2000)\citenamefont{Yuan, Kononov,
  Cavalcante, Lutchen, Ingenito, and Suki}}]{Yuan2000}
\bibinfo{author}{\bibfnamefont{H.}~\bibnamefont{Yuan}},
  \bibinfo{author}{\bibfnamefont{S.}~\bibnamefont{Kononov}},
  \bibinfo{author}{\bibfnamefont{F.~S.} \bibnamefont{Cavalcante}},
  \bibinfo{author}{\bibfnamefont{K.~R.} \bibnamefont{Lutchen}},
  \bibinfo{author}{\bibfnamefont{E.~P.} \bibnamefont{Ingenito}},
  \bibnamefont{and} \bibinfo{author}{\bibfnamefont{B.}~\bibnamefont{Suki}},
  \bibinfo{journal}{Journal of applied physiology (Bethesda, Md. : 1985)}
  \textbf{\bibinfo{volume}{89}}, \bibinfo{pages}{3} (\bibinfo{year}{2000}),
  ISSN \bibinfo{issn}{8750-7587}.

\bibitem[{\citenamefont{Chen et~al.}(2004)\citenamefont{Chen, Suki, and
  An}}]{Chen2004}
\bibinfo{author}{\bibfnamefont{Q.}~\bibnamefont{Chen}},
  \bibinfo{author}{\bibfnamefont{B.}~\bibnamefont{Suki}}, \bibnamefont{and}
  \bibinfo{author}{\bibfnamefont{K.-N.} \bibnamefont{An}},
  \bibinfo{journal}{Journal of biomechanical engineering}
  \textbf{\bibinfo{volume}{126}}, \bibinfo{pages}{666} (\bibinfo{year}{2004}),
  ISSN \bibinfo{issn}{01480731}.

\bibitem[{\citenamefont{Djordjevi\'{c}
  et~al.}(2003)\citenamefont{Djordjevi\'{c}, Jari\'{c}, Fabry, Fredberg, and
  Stamenovi\'{c}}}]{Djordjevic2003}
\bibinfo{author}{\bibfnamefont{V.~D.} \bibnamefont{Djordjevi\'{c}}},
  \bibinfo{author}{\bibfnamefont{J.}~\bibnamefont{Jari\'{c}}},
  \bibinfo{author}{\bibfnamefont{B.}~\bibnamefont{Fabry}},
  \bibinfo{author}{\bibfnamefont{J.~J.} \bibnamefont{Fredberg}},
  \bibnamefont{and}
  \bibinfo{author}{\bibfnamefont{D.}~\bibnamefont{Stamenovi\'{c}}},
  \bibinfo{journal}{Annals of Biomedical Engineering}
  \textbf{\bibinfo{volume}{31}}, \bibinfo{pages}{692} (\bibinfo{year}{2003}),
  ISSN \bibinfo{issn}{00906964}.

\bibitem[{\citenamefont{Duenwald et~al.}(2009)\citenamefont{Duenwald, Vanderby,
  and Lakes}}]{Duenwald2009}
\bibinfo{author}{\bibfnamefont{S.~E.} \bibnamefont{Duenwald}},
  \bibinfo{author}{\bibfnamefont{R.}~\bibnamefont{Vanderby}}, \bibnamefont{and}
  \bibinfo{author}{\bibfnamefont{R.~S.} \bibnamefont{Lakes}},
  \bibinfo{journal}{Annals of Biomedical Engineering}
  \textbf{\bibinfo{volume}{37}}, \bibinfo{pages}{1131} (\bibinfo{year}{2009}),
  ISSN \bibinfo{issn}{00906964}.

\bibitem[{\citenamefont{Soczkiewicz}(2002)}]{Soczkiewicz2002}
\bibinfo{author}{\bibfnamefont{E.}~\bibnamefont{Soczkiewicz}},
  \emph{\bibinfo{title}{{Application of Fractional Calculus in the Theory of
  Viscoelasticity}}} (\bibinfo{year}{2002}).

\bibitem[{\citenamefont{Fabry et~al.}(2003)\citenamefont{Fabry, Maksym, Butler,
  Glogauer, Navajas, Taback, Millet, and Fredberg}}]{Fabry2003}
\bibinfo{author}{\bibfnamefont{B.}~\bibnamefont{Fabry}},
  \bibinfo{author}{\bibfnamefont{G.~N.} \bibnamefont{Maksym}},
  \bibinfo{author}{\bibfnamefont{J.~P.} \bibnamefont{Butler}},
  \bibinfo{author}{\bibfnamefont{M.}~\bibnamefont{Glogauer}},
  \bibinfo{author}{\bibfnamefont{D.}~\bibnamefont{Navajas}},
  \bibinfo{author}{\bibfnamefont{N.~a.} \bibnamefont{Taback}},
  \bibinfo{author}{\bibfnamefont{E.~J.} \bibnamefont{Millet}},
  \bibnamefont{and} \bibinfo{author}{\bibfnamefont{J.~J.}
  \bibnamefont{Fredberg}}, \bibinfo{journal}{Physical review. E, Statistical,
  nonlinear, and soft matter physics} \textbf{\bibinfo{volume}{68}},
  \bibinfo{pages}{041914} (\bibinfo{year}{2003}), ISSN
  \bibinfo{issn}{1063-651X}.

\bibitem[{\citenamefont{Schiessel and
  Blumen}(1995{\natexlab{b}})}]{Schiessel1995Fractal}
\bibinfo{author}{\bibfnamefont{H.}~\bibnamefont{Schiessel}} \bibnamefont{and}
  \bibinfo{author}{\bibfnamefont{a.}~\bibnamefont{Blumen}},
  \bibinfo{journal}{Fractals} \textbf{\bibinfo{volume}{03}},
  \bibinfo{pages}{483} (\bibinfo{year}{1995}{\natexlab{b}}), ISSN
  \bibinfo{issn}{0218-348X}.

\bibitem[{\citenamefont{Schiessel et~al.}(1999)\citenamefont{Schiessel,
  Metzler, Blumen, and Nonnenmacher}}]{Schiessel1999}
\bibinfo{author}{\bibfnamefont{H.}~\bibnamefont{Schiessel}},
  \bibinfo{author}{\bibfnamefont{R.}~\bibnamefont{Metzler}},
  \bibinfo{author}{\bibfnamefont{a.}~\bibnamefont{Blumen}}, \bibnamefont{and}
  \bibinfo{author}{\bibfnamefont{T.~F.} \bibnamefont{Nonnenmacher}},
  \bibinfo{journal}{Journal of Physics A: Mathematical and General}
  \textbf{\bibinfo{volume}{28}}, \bibinfo{pages}{6567} (\bibinfo{year}{1999}),
  ISSN \bibinfo{issn}{0305-4470}.

\bibitem[{\citenamefont{Kelly and McGough}(2009)}]{Kelly2009}
\bibinfo{author}{\bibfnamefont{J.~F.} \bibnamefont{Kelly}} \bibnamefont{and}
  \bibinfo{author}{\bibfnamefont{R.~J.} \bibnamefont{McGough}},
  \bibinfo{journal}{The Journal of the Acoustical Society of America}
  \textbf{\bibinfo{volume}{126}}, \bibinfo{pages}{2072} (\bibinfo{year}{2009}),
  ISSN \bibinfo{issn}{00014966}.

\bibitem[{\citenamefont{Kurzweil}(2004)}]{kurzweil2004law}
\bibinfo{author}{\bibfnamefont{R.}~\bibnamefont{Kurzweil}},
  \emph{\bibinfo{title}{The law of accelerating returns}}
  (\bibinfo{publisher}{Springer}, \bibinfo{year}{2004}).

\bibitem[{\citenamefont{Mizuno et~al.}(2002)\citenamefont{Mizuno, Takayasu, and
  Takayasu}}]{mizuno2002mechanism}
\bibinfo{author}{\bibfnamefont{T.}~\bibnamefont{Mizuno}},
  \bibinfo{author}{\bibfnamefont{M.}~\bibnamefont{Takayasu}}, \bibnamefont{and}
  \bibinfo{author}{\bibfnamefont{H.}~\bibnamefont{Takayasu}},
  \bibinfo{journal}{Physica A: Statistical Mechanics and its Applications}
  \textbf{\bibinfo{volume}{308}}, \bibinfo{pages}{411} (\bibinfo{year}{2002}).

\bibitem[{\citenamefont{Hoshi et~al.}(2008)\citenamefont{Hoshi, Kobayashi, and
  Fujie}}]{Hoshi2008}
\bibinfo{author}{\bibfnamefont{T.}~\bibnamefont{Hoshi}},
  \bibinfo{author}{\bibfnamefont{Y.}~\bibnamefont{Kobayashi}},
  \bibnamefont{and} \bibinfo{author}{\bibfnamefont{M.~G.} \bibnamefont{Fujie}},
  \bibinfo{journal}{Proceedings of the 2nd Biennial IEEE/RAS-EMBS International
  Conference on Biomedical Robotics and Biomechatronics, BioRob 2008} pp.
  \bibinfo{pages}{730--735} (\bibinfo{year}{2008}).

\bibitem[{\citenamefont{Ma and Hori}(2004)}]{Ma2004}
\bibinfo{author}{\bibfnamefont{C.}~\bibnamefont{Ma}} \bibnamefont{and}
  \bibinfo{author}{\bibfnamefont{Y.}~\bibnamefont{Hori}},
  \bibinfo{journal}{Nonlinear Dynamics} \textbf{\bibinfo{volume}{38}},
  \bibinfo{pages}{171} (\bibinfo{year}{2004}), ISSN \bibinfo{issn}{0924090X}.

\end{thebibliography}

\end{document}